\documentclass[12pt,preprint]{aastex}
\usepackage{amsmath, mathrsfs}
\usepackage{fixltx2e}

\makeatletter
\renewcommand{\@make@caption@text}[2]{%
  \begin{center}
    \makebox[\textwidth]{\rmfamily#1.\quad#2}
  \end{center}
}%
\makeatother

\begin{document}

%\raggedbottom
%\renewcommand{\thesection}{\Roman{section}} % convert section numbers to roman
%\renewcommand{\thesubsection}{\Roman{subsection}} % convert section numbers to roman
\newcommand{\sgn}{\operatorname{sgn}}
\newcommand{\kepler}{{\it Kepler}}
\newcommand{\keplermission}{{\it Kepler} Mission}
\newcommand{\rearth}{\ensuremath{R_{\oplus}}}
\newcommand{\aearth}{\ensuremath{a_{\oplus}}}
\newcommand{\rsun}{\ensuremath{R_\sun}}
\newcommand{\tsun}{\ensuremath{T_\sun}}
\newcommand{\slfrac}[2]{\left.#1\middle/#2\right.} % allows for automatic sizing of slashes in fractions
\newcommand{\chitwotwo}{\ensuremath{\chi^2_{\left(2\right)}}}
\newcommand{\chitwogof}{\ensuremath{\chi^2_{\left(GOF\right)}}}
%\newcolumntype{C}[1]{>{\centering}m{#1}}

\title{Detection of Potential Transit Signals in 17~Quarters \linebreak of \textit{Kepler} Data: Results of the Final \textit{Kepler} Mission  \linebreak Transiting Planet Search (DR25)}

\author{Joseph D. Twicken\altaffilmark{1}, Jon M. Jenkins\altaffilmark{2}, Shawn E. Seader\altaffilmark{1}, Peter Tenenbaum\altaffilmark{1}, Jeffrey C. Smith\altaffilmark{1}, Lee S. Brownston\altaffilmark{3}, Christopher J. Burke\altaffilmark{1}, Joseph H. Catanzarite\altaffilmark{1}, Bruce D. Clarke\altaffilmark{1}, Miles T. Cote\altaffilmark{2}, Forrest R. Girouard\altaffilmark{4}, Todd C. Klaus\altaffilmark{3}, Jie Li\altaffilmark{1}, Sean D. McCauliff\altaffilmark{5}, Robert L. Morris\altaffilmark{1}, Bill Wohler\altaffilmark{1}, Jennifer R. Campbell\altaffilmark{5}, Akm Kamal Uddin\altaffilmark{5}, Khadeejah A. Zamudio\altaffilmark{5}, Anima Sabale\altaffilmark{5}, Steven T. Bryson\altaffilmark{2}, Douglas A. Caldwell\altaffilmark{1}, Jessie L. Christiansen\altaffilmark{6}, Jeffrey L. Coughlin\altaffilmark{1}, Michael R. Haas\altaffilmark{2}, Christopher E. Henze\altaffilmark{2}, Dwight T. Sanderfer\altaffilmark{2}, Susan E. Thompson\altaffilmark{1}} 
\altaffiltext{1}{SETI Institute/NASA Ames Research Center, Moffett Field, CA 94035, USA}
\altaffiltext{2}{NASA Ames Research Center, Moffett Field, CA 94035, USA}
\altaffiltext{3}{Stinger-Ghaffarian Technologies/NASA Ames Research Center, Moffett Field, CA 94035, USA}
\altaffiltext{4}{Logyx LLC/NASA Ames Research Center, Moffett Field, CA 94035, USA}
\altaffiltext{5}{Wyle Laboratories, Inc./NASA Ames Research Center, Moffett Field, CA 94035, USA}
\altaffiltext{6}{NASA Exoplanet Science Institute, California Institute of Technology, Pasadena, CA 91106, USA}
\email{joseph.twicken@nasa.gov}


%\twocolumn[
%\maketitle
%\begin{onecolabstract}
\begin{abstract}
We present results of the final \kepler{} Data Processing Pipeline search for transiting planet signals in the full 17-quarter primary mission data set. The search includes a total of 198,709 stellar targets, of which 112,046 were observed in all 17~quarters and 86,663 in fewer than 17~quarters. We report on 17,230 targets for which at least one transit signature is identified that meets the specified detection criteria: periodicity, minimum of three observed transit events, detection statistic (i.e.,~signal-to-noise ratio) in excess of the search threshold, and passing grade on three statistical transit consistency tests. Light curves for which a transit signal is identified are iteratively searched for additional signatures after a limb-darkened transiting planet model is fitted to the data and transit events are removed. The search for additional planets adds 16,802 transit signals for a total of 34,032; this far exceeds the number of transit signatures identified in prior pipeline runs. There was a strategic emphasis on completeness over reliability for the final \kepler{} transit search. A comparison of the transit signals against a set of 3402 well-established, high-quality \kepler{} Objects of Interest yields a recovery rate of 99.8\%. The high recovery rate must be weighed against a large number of false alarm detections. We examine characteristics of the planet population implied by the transiting planet model fits with an emphasis on detections that would represent small planets orbiting in the habitable zone of their host stars.
%\vspace{36pt}
%\end{onecolabstract}]
\end{abstract}

\section{INTRODUCTION}
\label{s:intro}
The results of past \keplermission{} transiting planet searches have been presented in \citet{pt2012} for Quarter~1 through Quarter~3 (i.e., Q1--Q3), \citet{pt2013} for Q1--Q12, \citet{pt2014} for Q1--Q16, and \citet{seader2015} for Q1--Q17. We now present results of the final \kepler{} transiting planet search encompassing the complete 17-quarter primary mission. The search culminated end-to-end reprocessing of the primary mission data set with the latest \kepler{} Data Processing Pipeline \citep{jenkins2010a} code base (SOC 9.3). Differences from the previous 17-quarter search are due to improvements in pipeline pixel calibration, photometry and transiting planet search algorithms, modifications to transit signal consistency tests and test criteria, and fixes for software code bugs.

The data release for the final Q1--Q17 pipeline processing is referred to as Data Release 25 (DR25). We shall employ the name DR25 to refer to the final pipeline transit search and final search results. Likewise, we shall refer to the prior Q1--Q17 transit search and associated results reported by \citet{seader2015} as DR24. 

The Q1--Q3 transit search results were not vetted for a specific \keplermission{} planetary candidates catalog. The Q1--Q12, Q1--Q16, and Q1--Q17 DR24 transit search results were vetted and published in catalogs of planetary candidates by \citet{rowe2015}, \citet{mullally2015}, and \citet{coughlin2016}, respectively. Additional \keplermission{} planetary candidate catalogs have also been published by \citet{borucki2,borucki3}, \citet{batalha1}, and \citet{burke1}.

Primary mission data acquisition was terminated when the second of the four \textit{Kepler} reaction wheels failed five days into the second month of Q17. The spacecraft was subsequently repurposed to observe targets in fields of view centered on the ecliptic plane in three-month campaigns as part of the \textit{K2 Mission} \citep{howell, vancleve}.

\subsection{\textit{Kepler} Science Data}
\label{ss:data}
The details of \kepler{} operations and data acquisition have been reported elsewhere \citep{science-ops}. In brief, the \kepler{} spacecraft is in an Earth-trailing heliocentric orbit and maintained a boresight pointing centered on $\alpha = 19^{\rm h}22^{\rm m}40^{\rm s}, \delta = +44\degr.5$ during the primary mission. The \kepler{} photometer acquired data on a 115-square-degree region of the sky.  The data were acquired on 29.4-minute intervals, colloquially known as ``long cadences.''  Long-cadence pixel values were obtained by accumulating 270 consecutive 6.02~s exposures. The spacecraft was rotated about its boresight axis by $90\degr$ every ${\sim}93$~days in order to maintain the correct orientation of its solar panels and thermal radiator; the time period corresponding to a particular rotation state is known as a ``quarter.'' Target stars were observed throughout the year in four different locations on the focal plane due to the quarterly rotation.

Science data acquisition was interrupted periodically: monthly for data downlink, quarterly for maneuvering to a new roll orientation (this was typically combined with a monthly downlink to limit the loss of observation time), and once every three days for reaction wheel desaturation (one long cadence interval was sacrificed at each desaturation). Data acquisition was also interrupted on an irregular basis due to spacecraft anomalies. In addition to these interruptions required for normal operation, data acquisition was suspended for 11.3~days in Q16, from 2013-01-17 19:39Z\footnote{Time and date are presented here in ISO-8601 format, YYYY-MM-DD HH:MM, or optionally YYYY-MM-DD HH:MM:SS, with a trailing `Z' to denote UTC.} through 2013-01-29 03:50Z (555 long-cadence samples). During this time, the spacecraft reaction wheels were commanded to halt motion in an effort to relubricate and thereby extend the life of the failing reaction wheel~4. Spacecraft operation without the use of reaction wheels is not compatible with high-precision photometric data acquisition, so science data were not collected during the wheel rest.

In 2012 July, one of the four reaction wheels employed to maintain spacecraft pointing during science acquisition experienced a catastrophic failure. The mission was able to continue using the remaining three wheels to achieve three-axis control of the spacecraft until 2013 May. At that time, a second reaction wheel failed, forcing an end to data acquisition in the nominal \kepler{} field of view. The results reported here represent the final  \kepler{} pipeline processing incorporating the full volume of data acquired from that field of view.\footnote{We exclude 10~days of data (dubbed Q0) acquired at the end of commissioning on ${\sim}53,000$ stars as this segment of data is too short to avoid undesirable edge effects in the transit search. A planetary candidates catalog based on results of a transit search of Q0--Q5 was published by \citet{borucki3}.}

Science acquisition of Q1 data began at 2009-05-13 00:01:07Z, and acquisition of Q17 data concluded at 2013-05-11 12:16:22Z. This time period contains 71,427 long-cadence intervals.  Of these, 5077 were consumed by the interruptions listed above.  An additional 1145 long-cadence intervals were excluded from searches for transiting planets because of data anomalies that came to light during processing and inspection of flight data.  These include a contiguous set of 255 long-cadence samples acquired over 5.2~days that immediately preceded the Q16 downtime described above. The shortness of this data set combined with the duration of the subsequent gap led to a judgment that the data would not be useful for transiting planet searches. The excluded intervals also include 372 long cadences (7.6~days) acquired during periods of coronal mass ejections in Q12; data quality at that time was not consistent with that maintained over the course of the primary mission. The remaining 65,205 long cadence intervals were employed in the Q1--Q17 transiting planet search.

A total of 198,709 targets observed by \kepler{} were searched for evidence of transiting planets in the DR25 pipeline run. The target set included all stellar targets observed by \kepler{} at any point during the mission and specifically included target stars that were not originally observed for the purpose of transiting planet search (asteroseismology targets, guest observer targets, and so on). There were two exceptions, however.  A subset of known eclipsing binaries was excluded from the transit search, as described below. In addition, 89 short-cadence (i.e.,~one-minute interval) guest observer targets that were also acquired at long cadence were inadvertently left off the transit search target list because of a subtlety of \keplermission{} target management.

Seven of the targets searched for transiting planets were designated as Custom Targets. Such targets are not stellar objects in the \textit{Kepler} Input Catalog \citep{brown}, but rather represent regions of pixels in given quarters that were collected in support of a variety of scientific investigations. Custom Targets were assigned numerical identifiers outside the range of the \textit{Kepler} Input Catalog (also known as the KIC). There was no attempt in the pipeline to compute optimal photometric apertures for Custom Targets; all pixels in the regions associated with each of these targets were employed to formulate the respective photometric apertures. A threshold crossing event (TCE) was identified for one Custom Target (KIC 100001645) based on observations in Q4, Q8, and Q12. TCEs represent potential transit signals that meet specified detection criteria and pass all statistical transit consistency tests.

\begin{figure}
%%\epsscale{.80}
%\plotone{fig1.eps}
% For ArXiv
\plotone{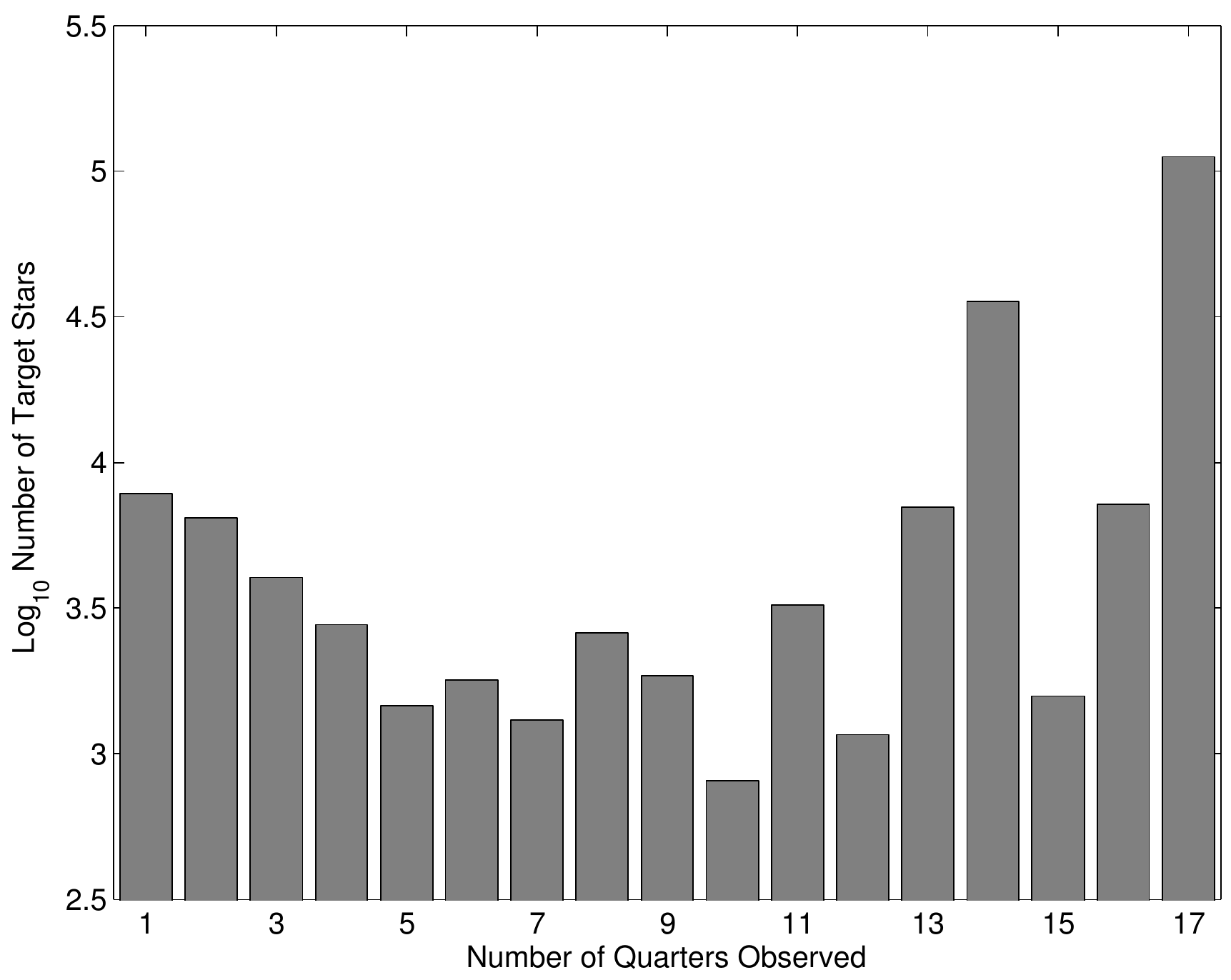}
\caption{Distribution of transit search targets by number of quarters observed. The number of targets is displayed on a logarithmic scale. In total, 198,709 targets were searched for transit signals.
\label{f1}}
\end{figure}

Figure~\ref{f1} shows the distribution of targets by the number of quarters of observation. The number of targets is displayed on a logarithmic scale. A total of 112,046 targets were observed for all 17~quarters. An additional 35,650 targets were observed for 14~quarters; the vast majority of these targets are in regions of the sky observed in some quarters by CCD Module~3, which experienced a hardware failure in its readout electronics in 2010 January during Q4. This resulted in a ``blind spot'' that rotated along with the \kepler{} spacecraft, effectively removing 25\% of the quarterly observations for affected targets.  The balance of 51,013 targets observed for some other number of quarters is largely due to changes in the target set over the duration of the mission. Table~\ref{t1} lists the KIC identifiers, number of quarters observed, and number of TCEs (see Section~\ref{ss:multiples}) associated with each of the targets that were included in the DR25 transit search; the table is sorted by KIC ID. 

Some known eclipsing binaries have been excluded from planet searches in the pipeline, as described in \citet{pt2014} and \citet{seader2015}. The pipeline transiting planet search and data validation algorithms are not well suited to binaries without detached and well-separated eclipses. A total of 975 known contact and semidetached eclipsing binaries were excluded from the Q1 to Q17 DR25 search. This is fewer than the number excluded in the earlier analyses (1519 in the Q1--Q16 search and 1033 in the Q1--Q17 DR24 search). We excluded eclipsing binaries from the \kepler{}  catalog of eclipsing binaries \citep{eb-cat1,eb-cat2,eb-cat3} as of 2015 July 1 that (1) have a morphology parameter $\ge$~0.6 \citep{morph} but (2) were not designated as \kepler{} Objects of Interest (KOIs) as of 2015 September 25 in the cumulative KOI table at the NASA Exoplanet Archive \citep{akeson}. There are 47 eclipsing binary KOIs in the DR25 analysis that would otherwise have been excluded by the morphology cut. The criteria for excluding eclipsing binaries from the transit search did not change from the DR24 run; the number of binaries that were excluded was reduced by changes in the eclipsing binary catalog and the cumulative KOI list. The KIC identifiers of the eclipsing binaries excluded from the DR25 transit search are listed in Table~\ref{t2}; this table is also sorted by KIC ID.

\subsection{Processing Sequence: Pixels to TCEs to KOIs}
\label{ss:sequence}
The steps in processing \kepler{} science data have not changed since \citet{pt2012} and are briefly summarized below. Improvements to pipeline pixel calibration and photometry algorithms between the DR24 and DR25 runs will be discussed in Section~\ref{ss:presearch}, and improvements to transit search algorithms will be discussed Section~\ref{s:tps}. 

The pixel data from the spacecraft were first calibrated, in the Calibration (CAL) pipeline module, to remove pixel-level effects such as bias, smear, gain variations, and nonlinearity \citep{calSpie2010}. The calibrated pixel values were combined after cosmic-ray removal and background subtraction in the Photometric Analysis (PA) pipeline module to produce a flux time series for each stellar target \citep{jdt2010, rlm2016}. The ensemble of target flux time series was then corrected in the Presearch Data Conditioning (PDC) pipeline module for systematic variations driven by effects such as differential velocity aberration, temperature-dependent focus changes, and instrument pointing excursions \citep{js2012,stumpe2012,stumpe2014}. The corrected flux time series produced by PDC were subsequently searched for transiting planet signals.

The Transiting Planet Search (TPS) software module analyzed each corrected flux time series individually for evidence of periodic reductions in flux that would indicate a possible transiting planet signature. The search process incorporated a significance threshold (7.1$\sigma$) against a multiple-event statistic (MES)\footnote{The multiple-event statistic is a measure of the degree to which the data are correlated with the reference waveform (in this case a sequence of evenly spaced transit pulses) normalized by the strength of the observation noise. It is approximately the same as the result of dividing the fitted transit depth by the uncertainty in the fitted transit depth and may be interpreted in terms of the likelihood a value would be seen at that level by chance.} and a series of statistical transit consistency tests (vetoes); the latter were necessary because the significance threshold was sufficient for rejection of the null hypothesis (i.e.,~no transit signal present) but was incapable of discriminating between multiple competing hypotheses, which can potentially explain the flux excursions. An ephemeris (period, epoch of first transit, and pulse duration) and associated MES on a given target that satisfied the significance threshold and passed all vetoes make up a TCE. The 7.1$\sigma$ transit search significance threshold was selected to yield on the order of one false alarm for a four-year \kepler{} mission given whitened Gaussian noise distributions and the numbers of stellar targets (200,000) and independent statistical tests ($10^{12}$) involved in the transiting planet search \citep{jenkins2002a}.

Each target with a TCE was searched for additional TCEs (through calls to the main TPS function) in the Data Validation (DV) software module; these potentially indicate multiple planets orbiting a single target star. Limb-darkened transiting planet models were fitted to the systematic error-corrected light curves in a whitened domain, and diagnostic tests were performed for each TCE in DV to assist in subsequent efforts to distinguish between true transiting planets and false alarm/false positive detections. The distinction between false alarms and false positives was discussed by \citet{bryson2016}. In principle, a TCE that has been accepted as a valid astrophysical signal (traced to either a transiting planet or eclipsing binary) is designated as a KOI.  A KOI that passes additional scrutiny is classified as Planet Candidate (PC), whereas one that does not is classified as False Positive (FP). \kepler{} TCE Review Team (TCERT) procedures for promoting TCEs to KOI status and classifying KOIs are discussed in \citet{rowe2015}, \citet{mullally2015,mullally2016}, \citet{coughlin2016}, and \citet{thompson2015a}. An alternative approach employing a machine learning system for automated vetting of TCEs identified in the pipeline has been described by \citet{mccauliff}, \citet{jenkins2014}, and \citet{catanzarite}. In this approach, a random forest of decision trees was trained to classify TCEs based on attributes (i.e.,~diagnostic metrics) produced by TPS and DV, and posterior probabilities were computed indicating the confidence in each individual classification.

\subsection{Pipeline Pixel Calibration and Photometry}
\label{ss:presearch}
There have been substantial improvements to the \kepler{} Data Processing Pipeline since the DR24 run with the SOC 9.2 code base. We describe here only these recent improvements and refer the interested reader to \citet{pt2012,pt2013,pt2014} and \citet{seader2015} for a description of past pipeline algorithms and enhancements.  There was a concerted effort to improve the photometric precision of the light curves and increase the sensitivity to small transiting planets with long orbital periods in the final  \kepler{} Pipeline code base. Photometric precision was improved for the bulk of nearly 200,000 target stars (a quantitative summary is presented in Section~\ref{sss:performance}), although there are individual targets for which the photometric precision of the SOC 9.3 light curves is inferior to that of earlier code bases and data releases. The \kepler{} Pipeline is fully automated, and the nature of the targets varies widely; one size has never fit all.

\subsubsection{Dynamic 2D Bias Correction}
The pixel-level calibration performed by the CAL module of the \kepler{} Pipeline was enhanced for the DR24 run to include a dynamic two-dimensional bias correction \citep{jeffk2010,clarke2012}; that is, a time dependent 2D bias correction was performed by module output and cadence (although not in Q0, Q1, and Q17). The dynamic correction accounts for drifts, temperature dependent behavior, and crosstalk in CCD electronics that the static correction employed earlier in the pipeline cannot. An issue was identified with the dynamic bias correction for pixels associated with long-cadence targets as first implemented in SOC 9.2 that was addressed in SOC 9.3. The dynamic bias correction accommodates step discontinuities across data gaps. In the SOC 9.2 code base, the step discontinuities were not applied on the correct cadences; the error was corrected in SOC 9.3. This improved the performance of the 2D bias correction for targets subjected to the DR25 transiting planet search, particularly targets for which there was significant temperature dependent behavior over the periods associated with data gaps.

\subsubsection{Rolling Band Artifacts}
In an effort to mitigate the effects of image artifacts described in \citet{caldwell}, rolling band artifact metrics \citep{clarke2015}  are now generated as a function of CCD row and cadence with discrete and floating-point severity levels for a set of five trial pulse durations (1.5, 3.0, 6.0, 12.0, and 15.5~hr). These metrics may be employed to identify the cadences on which individual targets were impacted by rolling band artifacts, and they are now available on the public archive. With SOC 9.3, the DV module also produces a rolling band contamination diagnostic for every TCE. This diagnostic identifies coincidences between transit events and rolling band artifacts at the trial pulse duration closest to the transit duration associated with the TCE \citep{jdt2014,jdt2016}. Although the rolling band contamination diagnostic does not impact whether or not a TCE is produced in the transit search, it may be employed when assessing the validity of each TCE. The image artifact metrics were produced only with discrete severity levels for a single pulse duration (10.5~hr) in the DR24 transit search.

\subsubsection{Photometric Apertures}
A major improvement in the SOC 9.3 code base involves the definition of photometric apertures. Prior to this final pipeline release, photometric apertures were optimized for photometric signal-to-noise ratio (S/N) based only on (1)~CCD electronics models,  (2)~pixel response function, and (3)~KIC magnitudes and coordinates for target stars and other objects in the field of view \citep{bryson2010}. For the DR25 run, photometric apertures were informed also by calibrated pixel data associated with each of the target stars in a given quarter. Pixel response functions were fitted jointly in PA to all catalog objects in the mask of pixels (32 pixels on average) associated with each target to better estimate the magnitude and positions of the targets and their celestial neighbors. Photometric apertures were then optimized for photometric precision in addition to S/N \citep{js2016}. Improved photometric precision translates directly to better sensitivity in the search for transiting planets. Given the revised, data-driven apertures, simple aperture photometry was performed to produce the DR25 light curves for target stars as before \citep{jdt2010, rlm2016}; the changes described here relate specifically to selection of the pixels that form the photometric aperture for each target and quarterly data set.

\subsubsection{Presearch Data Conditioning}
The PDC module was also enhanced in the latest pipeline release. Previous TCE populations have exhibited structure (lines) in orbital period versus epoch of first transit due to impulsive events on specific cadences that affect ensembles of target stars. Such structure can be seen in ``wedge'' plots of all TCEs in a given run (e.g.,~Figure~\ref{f4}). Hough transformations \citep{hough} have been utilized to project lines in wedge plots to common epochs and identify individual cadences that trigger excesses of TCEs. So-called ``spike'' basis vectors now fit impulsive flux excursions on these cadences separately from the cotrending basis vectors employed in PDC to remove systematic errors in target light curves \citep{stumpe2012}. Inclusion of the spike basis vectors in the systematic error removal process reduced the number of false alarms in the transit search due to common impulsive events.

PDC was also modified to perform Bayesian maximum a posteriori (MAP) systematic error correction \citep{js2012} within the shortest timescale band ($\leq 1.5$~hr) of the three into which the target light curves are split \citep{stumpe2014}. MAP cotrending was not previously applied to data at the shortest timescales. Doing so improved the correction of high-frequency/short-timescale systematics affecting the ensemble of target stars on each CCD module output in each quarter.

\subsubsection{Photometric Performance Summary}
\label{sss:performance}
The standard metric for characterizing \kepler{} photometric precision is combined differential photometric precision, or CDPP \citep{christiansen2012}. The overall improvement in pipeline photometric precision over that achieved in DR24 may be summarized as follows. For $Kp = 12$ targets (including dwarfs and giants), the DR25 pipeline rms CDPP was 30.44~ppm at the 10th percentile and 71.13~ppm at the 50th percentile for 6~hr integrations. This represents an improvement over DR24 of 1.65~ppm (5.1\%) at the 10th percentile and 7.21~ppm (9.2\%) at the 50th percentile. For $Kp = 15$ targets, the improvement in rms CDPP was 5.00~ppm (4.7\%) at the 10th percentile and 8.68~ppm (5.8\%) at the 50th percentile for 6~hr integrations.

All data were uniformly processed for the DR25 run with the SOC 9.3 code base on the NASA Advanced Supercomputing (NAS) Pleiades supercomputer cluster at NASA Ames Research Center. Pixel-level data and flux time series from these photometric processing steps are publicly available at the Mikulski Archive for Space Telescopes\footnote{http://archive.stsci.edu/index.html and http://archive.stsci.edu/kepler.} (MAST). This data release is documented at the archive in Data Release Notes 25 \citep{thompson2015b}. Rolling band artifact metrics by CCD row and cadence and by target and cadence (obtained from all rows intersecting the photometric aperture of each given target) are also available for the first time with the DR25 data products at MAST.

\section{TRANSITING PLANET SEARCH}
\label{s:tps}
This section describes changes that have been made to the TPS algorithms since the DR24 run.  For further information on the pipeline search for transiting planets, see \citet{jenkins2002b}, \citet{jmj2010b}, \citet{pt2012,pt2013,pt2014}, and \citet{seader2015}.

\subsection{Removal of Harmonic Content}
\label{ss:harmonics}
The adaptive, wavelet-based matched filter employed by TPS to search for transiting planet signals is not well suited to compact signals in the frequency domain, that is, sinusoidal signals such as those due to highly periodic pulsations. Therefore, harmonic signals are fitted and removed from each light curve on a quarter-by-quarter basis prior to conducting the transit search. This process significantly reduces the number of false alarms that would result from retaining these harmonic signatures; the process may also degrade or remove short-period ($<3$~days) transit signals \citep{christiansen2013,christiansen2015}. The harmonic fitting is conducted iteratively and is numerically intensive; a maximum number of harmonic components are therefore permitted to be fitted and removed in order to manage the time spent on this process and also to avoid overfitting.

Previous versions of the software allowed a fixed number of harmonics (25) to be removed from each quarterly light curve, regardless of the length of the quarter. This led to inconsistent fitting of periodic stellar variability wherein short quarters (e.g., Q1, Q4 for stars on CCD Module 3, and Q17) had significantly more harmonic content removed, artificially reducing the apparent observation noise and thereby biasing the transit search. The SOC 9.3 code base adjusts the maximum number of harmonic components fitted in each quarter to be proportional to the length of the quarterly data set, leading to more consistent harmonic fitting results across the full four-year data set. The maximum number of harmonic components removed per target and quarter in the DR25 run ranged from eight in Q17 and nine in Q1 at the low end, to 25 in Q9, Q11, Q14, and Q15.

\subsection{Whitening and Quarter Stitching}
\label{ss:whitening}
Several changes were made to the SOC 9.3 code base to improve the performance of the whitener and thereby increase the sensitivity of the transit search.

\subsubsection{Quarter-by-quarter Whitening}
Perhaps the most important change in TPS in SOC 9.3 is the introduction of quarter-by-quarter whitening. The previous code bases stitched all quarters of data together for each target star, then whitened the quarter-stitched flux time series and searched for transit-like features. The current version of TPS separately whitens the flux time series for each quarter prior to concatenating the quarterly segments together in preparation for the transit search. This change was motivated by the fact that a significant population of false alarms was generated near boundaries of the quarterly data segments. Inspection revealed that this was due to step changes that occurred in noise statistics and power from quarter to quarter as stars rotated onto new CCDs as a result of roll maneuvers. The adaptive, wavelet-based matched filter algorithm in TPS was designed to track gradual changes in the statistics of observation noise, not abrupt changes such as those engendered by roll maneuvers. Quarter-by-quarter whitening significantly reduced the incidence of false alarms near quarterly boundaries, as described further in Section~\ref{ss:matching}.

\subsubsection{Long-gap Filling}
Another important change to TPS in SOC 9.3 is a modification to the long-gap-fill algorithm. The new algorithm applies a narrow sigmoid taper for the periodic extension of time series across long data gaps ($>2.5$~days) prior to application of fast Fourier transforms for the wavelet filter bank. The sigmoid occupies the central 10\% of the gap to be filled; data are simply reflected from the left- and right-hand sides of the gap, then weighted by the sigmoid taper and added together in the central 10\% region. The former long-gap-fill algorithm used a linear taper across the entire gap, resulting in a systematic reduction in noise power estimates near the gap edges. This artificial drop in noise power was tolerable when the entire four-year time series was whitened at once, but it introduced significant biases in the sensitivity to transits or transit-like features near quarterly boundaries with quarter-by-quarter whitening. Modifying the long-gap-fill algorithm also improved the performance of TPS for long data gaps within quarterly data sets.

\subsubsection{Nondecimated Moving Median Absolute Deviation}
A third important improvement to TPS in SOC 9.3 is the incorporation of a nondecimated (in time) moving median absolute deviation (MAD) filter for estimating the rms noise power time series in each of the wavelet filter bank's bandpasses. This change was motivated by the observation that there was a measurable and significant duration-dependent bias in the noise power estimates: the noise power for short-duration transits was underestimated relative to that for long-duration transits. Prior to SOC 9.3, TPS employed a decimated moving MAD filter due to computational run time constraints; the moving MAD was not calculated for each sample in each of the wavelet bandpasses. We were able to implement the moving MAD filter algorithm more efficiently and perform the computation without decimation; this eliminated the bias in the noise power estimates while maintaining adequate computational throughput. 

These changes to TPS increase the sensitivity to transiting planet signatures, improve the uniformity of the sensitivity of the search, and enhance the performance and characteristics of the statistical bootstrap analysis \citep{jenkins2015} performed for each TCE in DV.

\subsection{False-alarm Vetoes}
\label{ss:vetoes}
As stated earlier, the adaptive matched filter implemented in TPS is augmented by additional statistical tests to determine whether a potential sequence of transit-like features is declared a TCE. A total of 68,004 target stars produced signatures with MES $> 7.1\sigma$ in the DR25 TPS run before application of the statistical filters. Light curves with signatures above $7.1\sigma$ that do not receive a passing grade on all statistical tests are searched further for lower MES signals until (1)~time is no longer available, (2)~an iteration limit is reached, (3)~no further signal with MES $> 7.1\sigma$ can be identified, or (4) a signal with MES $> 7.1\sigma$ is identified that survives all vetoes. The veto process was described in detail by \citet{seader2015}.

The robust statistic performs a kernel-based robust fit of the in-transit data to the putative transit signature and normalizes the fitted transit depth by the fit uncertainty. This test penalizes outliers that erroneously contribute to a high detection statistic (MES). The threshold for the robust statistic test was set to $6.8\sigma$ in the DR25 run and rejected 43,313 signatures with MES $>7.1\sigma$; that is, these signals failed the robust statistic test, and a lower MES signal was not subsequently identified that both met the detection threshold and passed all vetoes. The surviving 24,691 signatures were subjected to two different $\chi^2$ tests \citep{seader2013}. Thresholds for the robust statistic and $\chi^2$ tests were tuned by analysis of TPS performance for target stars with injected transit signatures and a subset of known KOIs; the thresholds were set as high as possible to facilitate false-alarm rejection without significantly impacting the recovery of true transit signals.

The \chitwotwo\ test breaks up the MES into different components, one for each transit event, and compares what is expected from each transit to what is actually obtained in the data assuming that there is indeed a transiting planet \citep{seader2013,seader2015}. In SOC 9.3, the calculation of the number of degrees of freedom for the \chitwotwo\ veto was updated to account for the use of nonrectangular, astrophysically motivated transit pulse shapes now employed in TPS. Monte Carlo experiments were conducted to determine an empirical correction factor~$\epsilon$ to the number of degrees of freedom for this veto as
\begin{equation}
\chitwotwo\left(DOF\right) = n_{Transits} - 1 + 
\frac{\epsilon \left(2 - \epsilon \right)}  {\left( 1-\epsilon \right)^2} Z^2,
%chiSquareResultsStruct.chiSquareDof2 = nTransits - 1 + epsilon * (2 - epsilon)/((1-epsilon)^2) * zCompSum^2 ;
\end{equation}
where $\epsilon \approx 0.04$ and $Z$ is defined as in Eq. 19 of \citet{seader2013}. The threshold for \chitwotwo\ in the DR25 run was set to $7.0\sigma$. The empirical correction factor was initially formulated by \citet{allen}.

The \chitwogof\ veto is similar to the \chitwotwo\ veto but involves the flux time series at the level of individual data points participating in the detection rather than at the level of transits. Thus, it has a higher number of degrees of freedom compared to the \chitwotwo\ veto. The threshold for \chitwogof\ was set to $6.8\sigma$ (reduced from $7.0\sigma$ in the DR24 run).

These two vetoes complement the MES and the robust statistic, which measure the correlation between the data and the fitted waveform. In contrast, \chitwotwo\ and \chitwogof\  measure the degree to which the fit residuals meet expectations, that is, the degree to which the scatter of the differences between the data and the fitted model is consistent with a zero-mean, unit-variance noise process as expected in the whitened domain. A total of 7461 of the potential DR25 TCEs that survived the robust statistic test failed at least one of the $\chi^2$ tests, and 5147 failed both tests; these represent the $\chi^2$ test failures after which lower MES TCEs were not subsequently identified. The number of potential TCEs that were rejected only by the \chitwotwo\ test was 1990 while the number rejected only by the \chitwogof\ test was 324. A total of 17,230 signatures with MES $> 7.1\sigma$ survived all statistical checks and were processed through DV;  this includes 1779 TCEs identified after stronger MES signals were vetoed.

\begin{figure}
%%\epsscale{.80}
%\plotone{fig2.eps}
\plotone{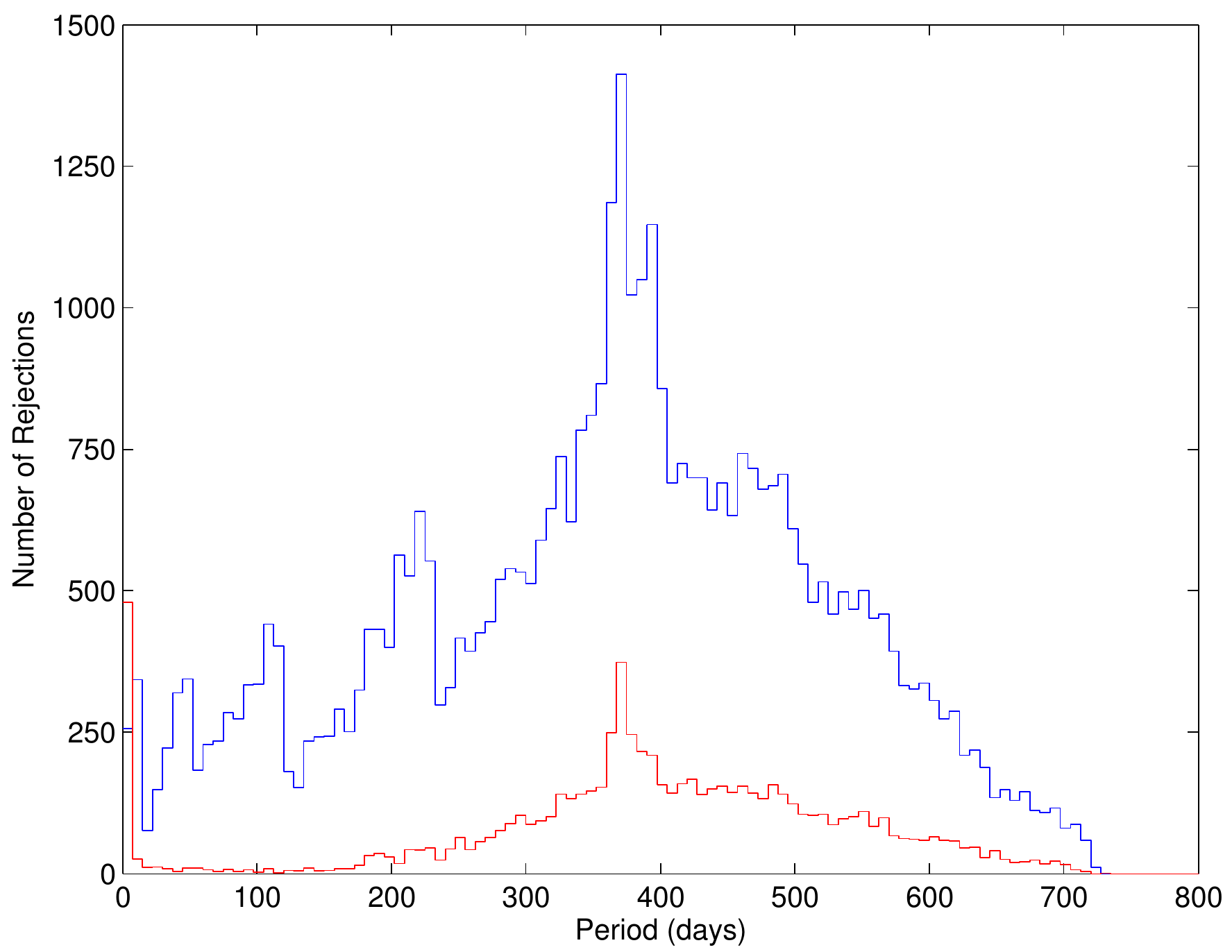}
\caption{Distribution of transit consistency vetoes by period in days. The number of (strongest MES) TCEs rejected by the robust statistic veto in each 7.5-day-period bin is displayed in blue. The number of (strongest MES) TCEs rejected by either of the $\chi^2$ vetoes in each period bin is displayed in red.
\label{f2}}
\end{figure}

The distribution of transit consistency vetoes in 7.5-day-period bins is displayed in Figure~\ref{f2}. Periods are shown in units of days on a linear scale. TCE rejections due to the robust statistic veto are displayed in blue, and rejections due to either of the $\chi^2$ vetoes are displayed in red. The $\chi^2$ vetoes were effective at eliminating TCEs at the shortest periods and for periods longer than 200~days. The distribution of the robust statistic vetoes is more uniform with period, although both types of vetoes exhibit strong peaks near the orbital period and hence thermal cycle of the \kepler\ spacecraft (372~days).

A fourth veto was in place for the Q1--Q17 DR24 run but was disabled for the DR25 run. This was a statistical bootstrap veto to enforce a uniform false-alarm probability across all TCEs. The bootstrap veto was described by \citet{seader2015}. Implications of disabling the bootstrap veto, improving the computation of the degrees of freedom associated with the \chitwotwo\ veto, and lowering the \chitwogof\ threshold for the DR25 TCE population will be discussed in Section~\ref{ss:multiples}.

\subsection{Incomplete Searches}
\label{ss:incomplete}
 \kepler{} Pipeline modules are subject to total processing time or ``wall time'' limits when run on the NAS Pleiades supercomputer. Computational tasks that are still running on Pleiades when the wall time limit is reached are killed; it is imperative that pipeline tasks not suffer this fate because results are not produced in such eventualities. The time limits for TPS and DV must accommodate transit searches and transit consistency checks, transiting planet model fitting, computation of diagnostic metrics, and generation of reports. The time limits are managed per target by enforcing limits on the number of TCEs, transiting planet search and consistency check iterations, and model fit and diagnostic test iterations. Furthermore, self-timeouts are enforced on a number of TPS and DV processing steps. It should be noted that whereas all TCEs produced in the pipeline are guaranteed to meet the pipeline detection threshold, there is no guarantee that a TCE represents the ``best'' detection (i.e.,~highest possible MES with a passing grade on all consistency tests) for the given light curve in the gridded search space. A TCE may represent the best result that could be achieved in the time available.
 
In the DR25 run, 35~hr were allocated for each TPS work unit compared to 50~hr for the DR24 run. The decision to do so was based on nontechnical considerations. Decreasing the time limit resulted in a small population of targets (1.3\% of the total) for which not all trial transit pulse durations were searched for transits.\footnote{Each light curve is searched in TPS for transits over 14 distinct pulse durations from 1.5 to 15~hr, starting from the longest pulse duration and working toward the shortest. If TPS reaches an internal timeout before the shortest pulse durations are searched, it will abort the search to preserve the results from the other pulse durations.}

Of 2497 targets that timed out before visiting all pulse durations, 1952 generated TCEs nonetheless. Indeed, only 545 of the targets that timed out failed to generate TCEs. We can determine whether significant sensitivity was lost due to the change in TPS time limit by investigating the fates of the 587 known KOIs for which not all pulse durations were searched in the DR25 run. Of these KOIs, 12 are confirmed or validated planets, 137 are classified as PC, and 438 are FP. TCEs were generated for 562 of these KOIs, most at the expected ephemeris. All 12 confirmed or validated planet light curves generated TCEs at the correct period and epoch. Only 25 of the 587 known KOIs failed to produce TCEs: four of these were PCs, and 21 were FPs. 

We conclude the following for the PCs that failed to produce TCEs based on DR24 transit search results: (1)~KOI 6262.01 is consistent with a transiting planet with an orbital period (0.34~days) below the minimum searched in the pipeline (0.5~days); (2)~KOI 7572.01 features purported transits with a period of 91.1~days that were largely gapped\footnote{Flux data on cadences in and near transit for a given TCE are removed (``gapped'') from the flux time series in DV before the search is invoked for additional transit signatures. The transits of subsequent, generally lower MES signatures that overlap the events associated with prior TCEs on the same target are therefore unavailable for the transit search algorithm and do not contribute to the detection MES. Short-period TCEs leave a periodic train of short data gaps in the residual flux time series searched for subsequent transit signals; these time series have been referred to as ``Swiss cheese.'' TCEs identified subsequent to short-period TCEs on a given target should be closely examined for validity.} by two short-period TCEs on the same target, and it does not appear to be a credible PC;  (3)~KOI 6918.01 appears likely to be an eclipsing binary and has been classified as such by \citet{eb-cat3}; (4)~KOI 6598.01 features purported transits with a period of 11.0~days that were largely gapped by three short-period TCEs triggered by strong stellar variability, and it also does not appear to be a credible PC. Thus, the self-timeouts affected none of the confirmed or validated planets, few credible PCs, and a small number of astrophysical FPs. While unfortunate, the incomplete transit searches in TPS are unlikely to significantly impact catalog completeness.

The search for additional planets in light curves with TCEs is conducted in DV (through internal calls to the main TPS function); the time allocated to these searches is also subject to run-time constraints. Sufficient processing time must be held in reserve so that the DV diagnostic tests may be performed and reports generated for all TCEs identified on a given target. The time limit on transit searches for additional planets in light curves with TCEs depends upon the total DV time allocation for each work unit (which did not change from DR24 to DR25) rather than the TPS allocation per work unit (which was reduced from DR24 to DR25). In the limit, the multiple-planet search is halted before any of the 14 trial pulse durations are searched if insufficient time would then be available to complete DV.

\subsection{Detection of Multiple-planet Systems}
\label{ss:multiples}
For the 17,230 target stars found to contain a TCE, additional transit searches were employed to identify potential multiple-planet systems. The process is described in \citet{hw2010}, \citet{pt2013}, and \citet{jdt2016}. The multiple planet search incorporates a configurable upper limit on the number of TCEs per target,  which is currently set to 10. This limit was established for two reasons. First, the limit on TCEs for a given target was instituted to manage pipeline task processing time as described earlier. Second, applying a limit to the number of TCEs per target prevents a failure mode in which a target flux time series is sufficiently pathological that the search process becomes ``stuck,'' returning one detection after another. The selected limit of 10~TCEs is based on experience: the maximum number of KOIs to date on a single target star is seven, which indicates that limiting the process to 10 TCEs per target is unlikely to sacrifice potential KOIs.

\begin{figure}
%%\epsscale{.80}
%\plotone{fig3.eps}
\plotone{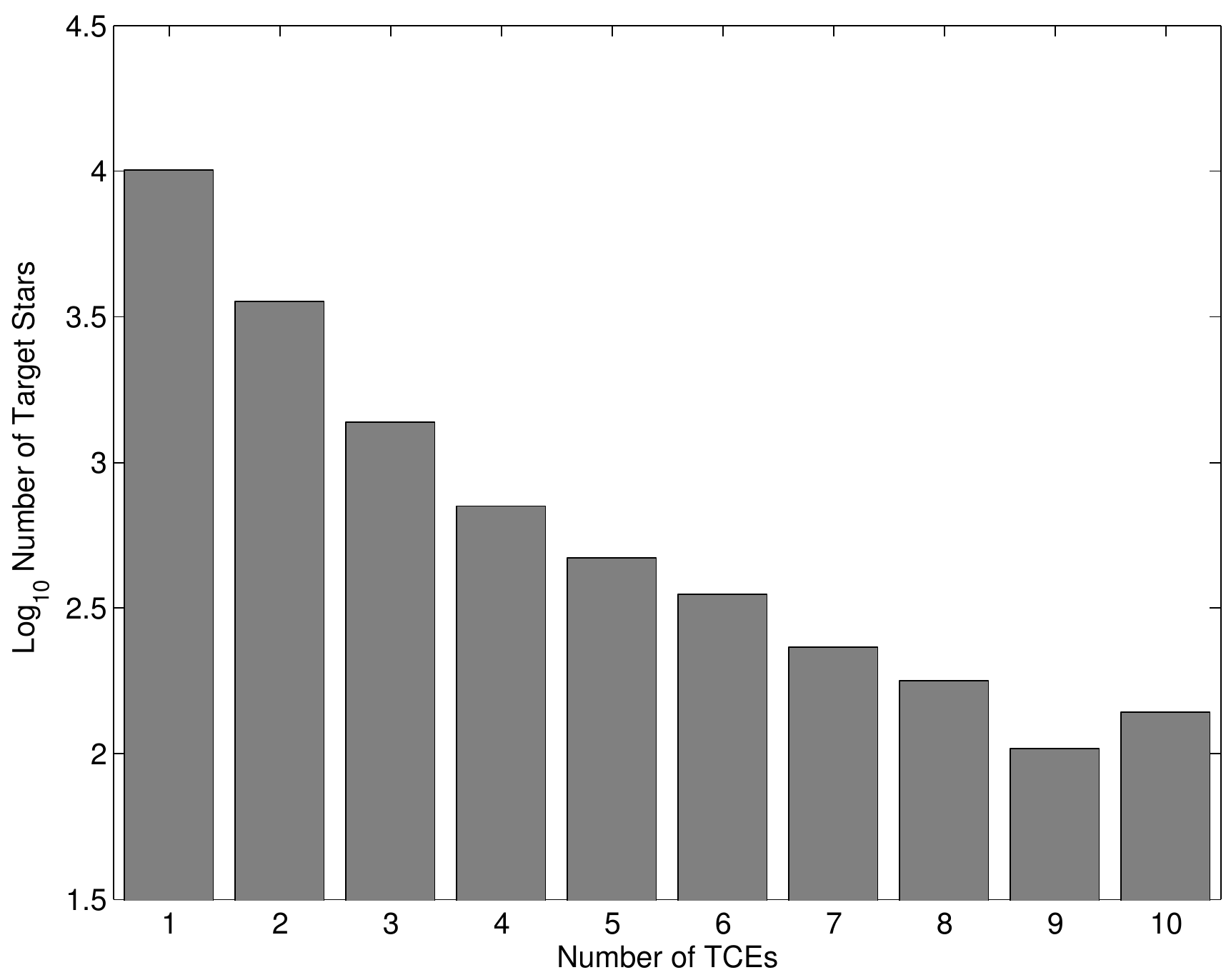}
\caption{Distribution of transit search targets by number of associated TCEs. The number of targets is displayed on a logarithmic scale. The maximum number of TCEs per target was configured to be 10. In total, 34,032 DR25 TCEs were identified on 17,230 unique targets for an average of 2.0 TCEs per target (for targets with TCEs).
\label{f3}}
\end{figure}

The transit searches performed for detection of TCEs in multiple-planet systems yielded 16,802 additional TCEs across 7120 unique target stars, for a total of 34,032 TCEs. The number of TCEs identified on each target in the DR25 transit search is listed in Table~\ref{t1}. The average number of TCEs per target with at least one potential transit signal is 2.0. Figure~\ref{f3} shows the distribution of the number of targets with each of the allowed numbers of TCEs; the number of targets is displayed on a logarithmic scale. The DR25 results represent the largest number of TCEs that have ever been produced in a pipeline run for generation of a catalog of planetary candidates. The \kepler{} Pipeline development team was encouraged by the science community to increase sensitivity to small planets in long-period orbits and to emphasize completeness over reliability. Accordingly, a large number of the 34,032 TCEs are false-alarm detections. The quality and reliability of the DR25 TCEs will be discussed in Section~\ref{ss:quality}. A variety of approaches may be undertaken to identify the subset of legitimate planet candidates from the large population of TCEs.

For the record, there were 7492 total TCEs in 1006 systems with six or more TCEs. The limit of 10 TCEs was reached for 139 targets. A small number of the TCEs in systems with six or more TCEs represent bona fide PCs and in some cases confirmed planets (e.g.,~KOI 157/Kepler-11 system, KOI 435/Kepler-154 system, KOI 351/Kepler-90 system). The vast majority of these TCEs are false alarms, however. There are not nearly this many detectable transiting planets in large multiple-planet systems in the \kepler{} data set. The bulk of these TCEs are triggered by artifacts and other features in individual light curves that produce multiple false alarms.

All TCEs included in this analysis have been delivered to the NASA Exoplanet Archive\footnote{http://exoplanetarchive.ipac.caltech.edu.} along with comprehensive DV Reports for each target with at least one TCE and one-page DV Report Summaries for each TCE. The reports and summaries are available to Exoplanet Archive visitors in PDF format. Tabulated DV model fit and diagnostic test results are also available at the Exoplanet Archive, as are newly redesigned data products in flexible image transport system (FITS) format that include TPS and DV time series data relevant to the TCEs identified in the pipeline \citep{thompson2016}.

\section{DETECTED SIGNALS OF POTENTIAL TRANSITING PLANETS}
\label{s:detected-signals}
The final search (DR25) for transiting planets in the full primary \keplermission{} data set produced 34,032 TCEs on 17,230 unique stellar targets. This compares with 16,285 TCEs on 9,743 unique targets in the Q1--Q16 search reported by \citet{pt2014}, and 20,367 TCEs on 12,669 unique targets in the Q1--Q17 DR24 search reported by \citet{seader2015}. The increase in the number of potential transit signatures is due largely to (1)~improvements in pipeline pixel calibration, photometry, and transiting planet search algorithms; (2)~modifications to transit signature consistency tests and test criteria; and (3)~software bug fixes. The fundamental transiting planet detection threshold has not changed, however.

\subsection{TCE Population}
\label{ss:tce-population} 
We now summarize the population of TCEs produced in the \kepler{} Data Processing Pipeline in the Q1--Q17 DR25 run and draw comparisons with the DR24 results reported by \citet{seader2015}.

\begin{figure}
%%\epsscale{.80}
%\plotone{fig4.eps}
\plotone{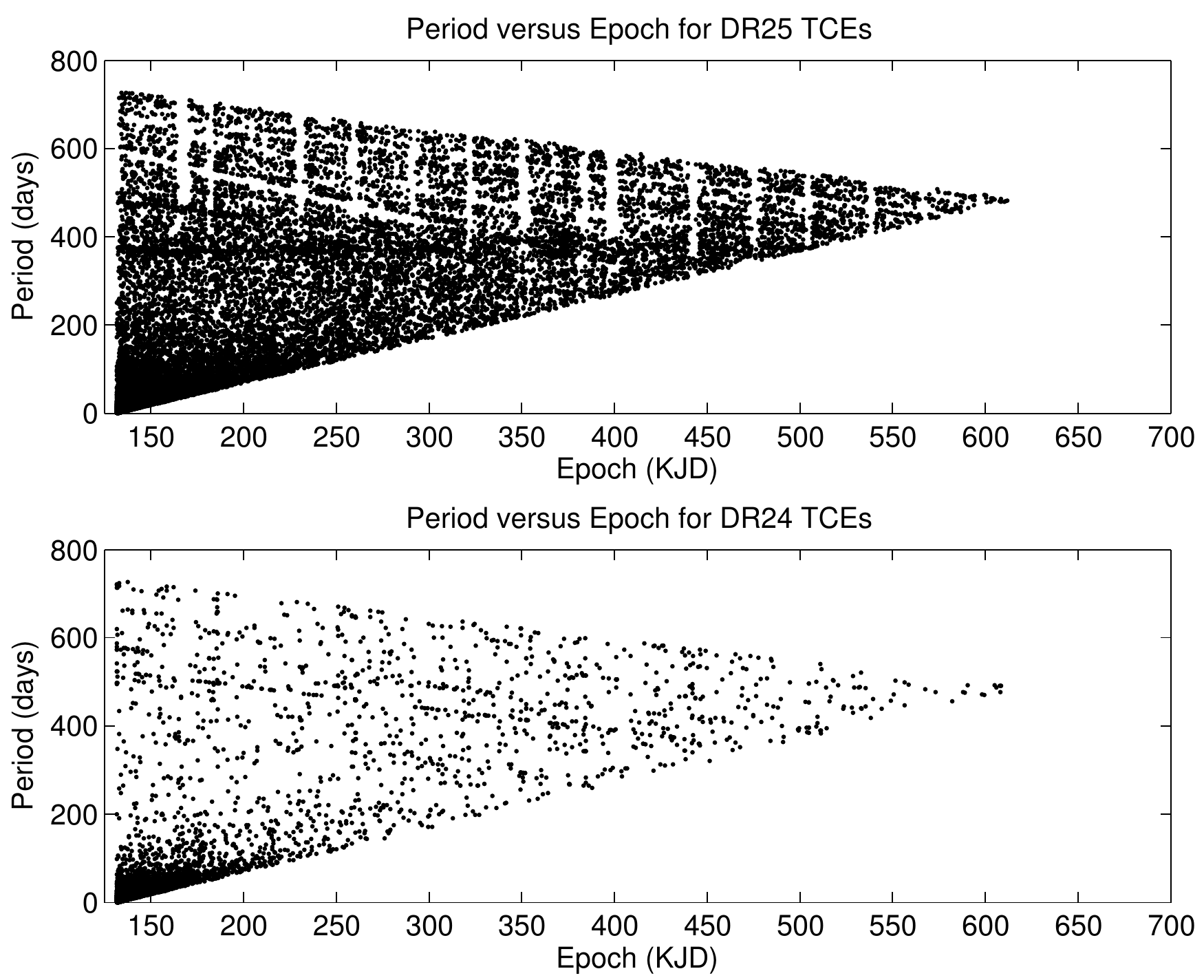}
\caption{Orbital period versus epoch ``wedge'' plots for the 34,032 DR25 TCEs detected in Q1--Q17 of \kepler{} data (top) and for the 
20,367 DR24 TCEs detected in Q1--Q17 of \kepler{} data (bottom) as reported in \citet{seader2015}. Periods are in days, and epochs are in \textit{Kepler}-modified Julian date (KJD); see text for definition.
\label{f4}}
\end{figure}

Figure~\ref{f4} (top panel) shows the period and epoch of first transit for each of the 34,032 DR25 TCEs, with period in units of days and epoch in \textit{Kepler}-modified Julian date (KJD), which is Julian date $-$ 2454833.0 (2009 January 1). As discussed earlier, structure in the ensemble of TCEs displayed in the period versus epoch ``wedge'' is undesirable (although white strips without TCEs are unavoidable, due to gaps in the \kepler{} data set). Figure~\ref{f4} (lower panel) shows the same plot for the 20,367 TCEs detected in the earlier DR24 pipeline run.  The axis scaling is identical for the two subplots, as is the marker size. Several features are apparent in this comparison.  First, the number of TCEs is considerably larger in the Q1--Q17 DR25 analysis. Second, the number of TCEs with long periods is considerably greater in the DR25 analysis.

The paucity of long-period TCEs in the prior analysis reported by \citet{seader2015} may be traced largely to the TPS transit signal consistency tests that were employed in the DR24 run. In an effort to mitigate the large number of long-period false alarms due to image artifacts and the quarterly photometer roll, the TPS vetoes eliminated many long-period TCEs. Preventing false alarms while at the same time maintaining sensitivity to transit signals has always been a difficult balancing act. As discussed earlier, we have (1)~improved the algorithm for computing the degrees of freedom associated with the \chitwotwo{} transit signal consistency test, (2)~lowered the threshold on the \chitwogof{} test, and (3)~disabled the consistency test based on a statistical bootstrap. Together with improvements to the quality of the light curves and the transiting planet search algorithm, we are hopeful that a number of the long-period TCEs in this Q1--Q17 analysis represent small planets orbiting in the habitable zone of Sun-like stars, even if the vast majority of the long-period TCEs represent false alarms.

\begin{figure}
%%\epsscale{.80}
%\plotone{fig5.eps}
\plotone{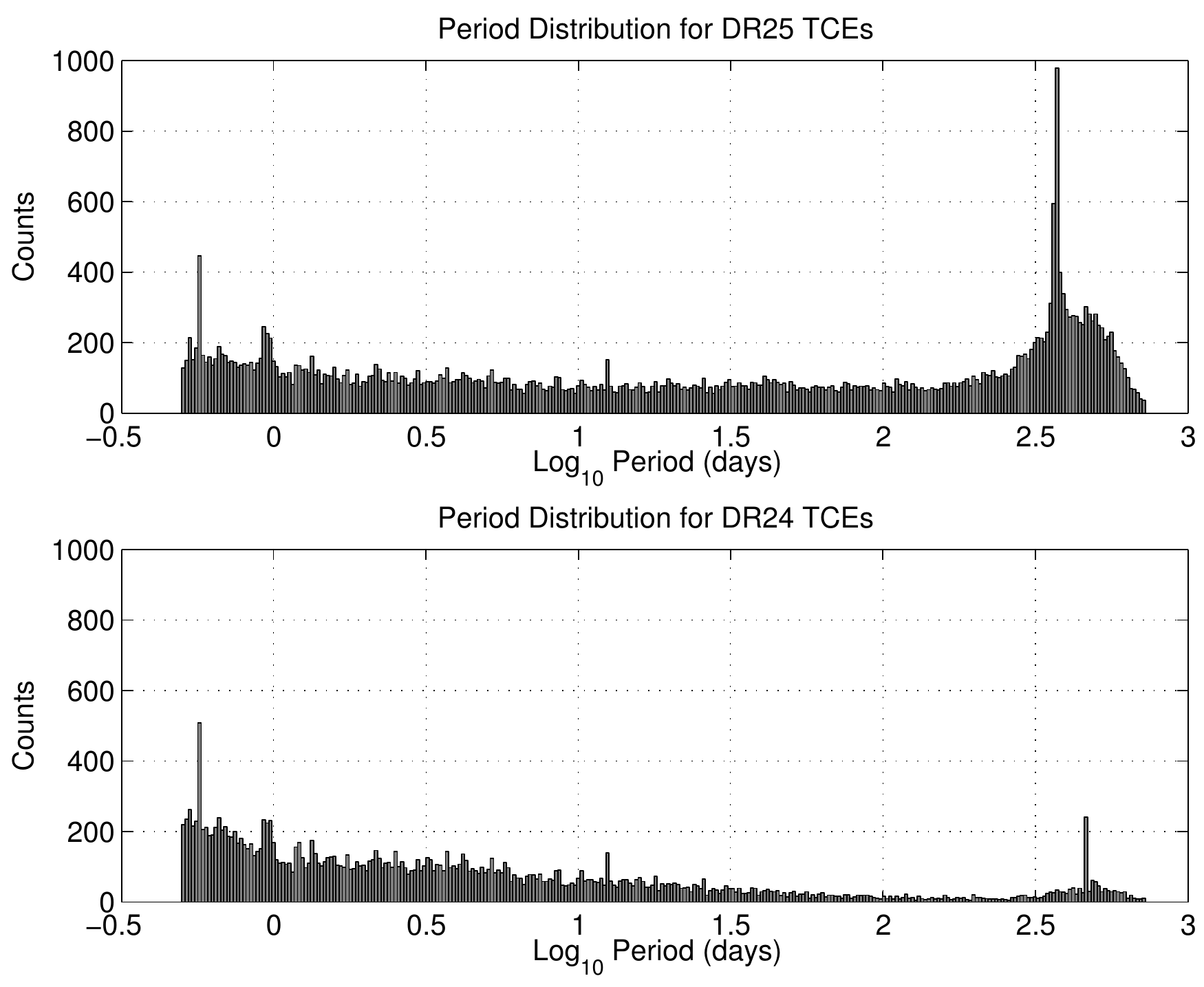}
\caption{Distribution of TCE periods plotted logarithmically. Top: 34,032 DR25 TCEs detected in Q1--Q17 of the \kepler{} data. Bottom: 
20,367 DR24 TCEs detected in Q1--Q17 of the \kepler{} data as reported in \citet{seader2015}. The peak near 372~days (log\textsubscript{10} = 2.57) in the DR25 results is coincident with the orbital period (and hence thermal cycle) of the \kepler{} spacecraft. The peak near 460~days (log\textsubscript{10} = 2.66) in the DR24 analysis has been eliminated by improvements to data-gap-filling code in the DR25 code base. The common peaks near 0.57~days (log\textsubscript{10} = -0.24) and 12.45~days (log\textsubscript{10} = 1.10) are due to contamination by RR Lyrae and V380 Cyg, respectively \citep{jcough}. 
\label{f5}}
\end{figure}

The drastic change in the distribution of TCE periods may be seen more clearly in Figure~\ref{f5}, which shows the distribution of TCE periods in units of days on a logarithmic scale. The Q1--Q17 DR25 results are shown in the upper panel of the figure, and the Q1--Q17 DR24 results are shown in the lower panel. The distributions of TCEs by orbital period are displayed with the same axis scaling in both cases. TPS has been configured to search for transiting planet signatures with orbital periods above 0.5~days. The minimum orbital period for the planet search was reduced from 1.0 to 0.5~days approximately 1.5~years after the \kepler{} launch to address the growing population of transiting planets with periods less than 1.0~day. The minimum orbital period was never reduced further in the pipeline search for transiting planets. There is a high computational cost involved in searching for very short period transit signatures. Short-period transiting planet signals do not necessarily go undetected, however; transit signals with periods below 0.5~days often produce one or more detections at integer multiples of the true orbital period. In cases such as these, the transit ephemerides are typically corrected by TCERT in the process of catalog generation. Failure to detect certain short-period transiting planets in the pipeline is likely to be attributable to removal of harmonic content (as discussed in Section~\ref{ss:harmonics}) or transit consistency vetoes (as discussed in Section~\ref{ss:vetoes}) rather than a lower limit on the transit search period.

The distribution of the latest Q1--Q17 TCEs is roughly uniform with log period up to periods on the order of 200 days. Beyond that point, there is a large excess of TCEs with a distinct peak near the \kepler{} orbital period of 372 days. This excess of long-period TCEs is clearly nonastrophysical. Transit signatures at long periods are composed of relatively few transits. Gaussian detection statistics do not necessarily apply when small numbers of transit events are folded; the false-alarm rate therefore depends not only on the pipeline detection threshold but also on the number of observed transits. Furthermore, the discriminating power of the $\chi^2$ consistency tests is diminished at long periods due to the low number of transit events and hence degrees of freedom. The false-alarm probability for each TCE is determined with a statistical bootstrap calculation in DV. The distribution of false alarm probabilities for the population of TCEs in the DR25 analysis will be discussed in Section~\ref{ss:quality}.

In the Q1--Q17 DR24 analysis reported by \citet{seader2015}, a bootstrap-based veto was employed in the suite of TPS transit signal consistency tests. This veto essentially enforced a detection threshold that varied by TCE to yield a uniform false-alarm probability. As seen in Figure~\ref{f5}, the distribution of TCEs generally decreased with (log) period, although there was also an excess at long periods. The increase in detection threshold with period reflected the non-Gaussian nature of the noise and was effective at eliminating long-period false alarms. In order to quantify the sensitivity and detection efficiency of the DR24 pipeline, transit signatures were injected at the pixel level into flight data associated with most long-cadence targets, and the \kepler{} Pipeline was subsequently run through PA, PDC, TPS, and DV. In the analysis of this run by \citet{christiansen2016a}, some loss in sensitivity of the DR24 pipeline to long-period transiting planets was noted. A similar transit injection activity is underway to characterize the DR25 \kepler{} Pipeline code base \citep{christiansen2016b}.

A close examination of Figure~\ref{f5} reveals that there are several peaks in the TCE period histograms for the DR24/DR25 results that are highly localized. There was a long-period peak near 460~days (log\textsubscript{10} = 2.66) in the DR24 analysis that has been eliminated by improvements to data-gap-filling code in SOC 9.3. There is a peak in the DR25 results near 372~days (log\textsubscript{10} = 2.57) that is due largely to image artifacts that repeat on the annual \kepler{} thermal cycle; this was also present in the Q1--Q16 results \citep{pt2014}. In both Q1--Q17 runs there are peaks attributable to bright, short-period sources in the \kepler{} field of view. The peaks near 0.57~days (log\textsubscript{10} = -0.24) and 12.45~days (log\textsubscript{10} = 1.10) are due to contamination by RR Lyrae and V380 Cyg, respectively \citep{jcough}. 

\begin{figure}
%%\epsscale{.80}
%\plotone{fig6.pdf}
\includegraphics[bb=38 14 510 394]{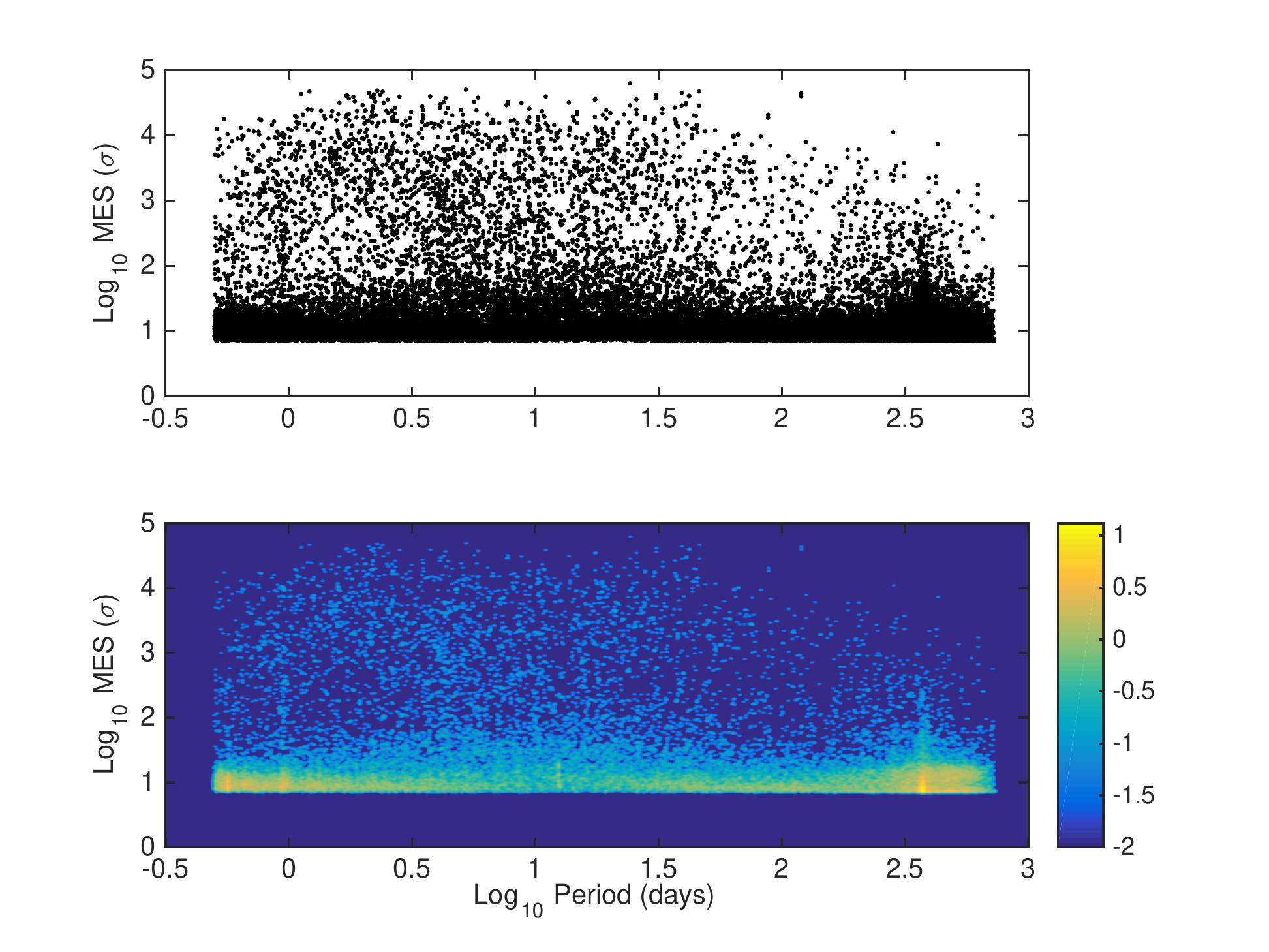}
\caption{Multiple-event statistic vs. orbital period in days on logarithmic scales. Top: all DR25 TCEs. Bottom: density for all TCEs on logarithmic scale. 
\label{f6}}
\end{figure}

Figure~\ref{f6} shows the TPS MES versus orbital period in days for the DR25 TCEs. The axes are displayed on logarithmic scales. All 34,032 TCEs are shown in the top panel. The bottom panel displays the density of the distribution. There is a dense band of TCEs associated with relatively low detection statistics across the full range of periods. There is an excess of long-period, low-MES TCEs that is dominated by false alarms. As discussed earlier, the vetoes are not as effective at long orbital periods with relatively few transit events. Furthermore, false-alarm probability increases for long-orbital periods because noise statistics based on relatively few transit events are not approximated well by a Gaussian distribution, and we did not increase the transit detection threshold to maintain a uniform false-alarm rate (as was effectively enforced with the bootstrap veto in DR24). We discuss this further in Section~\ref{ss:quality}.

\begin{figure}
%%\epsscale{.80}
%\plotone{fig7.eps}
\plotone{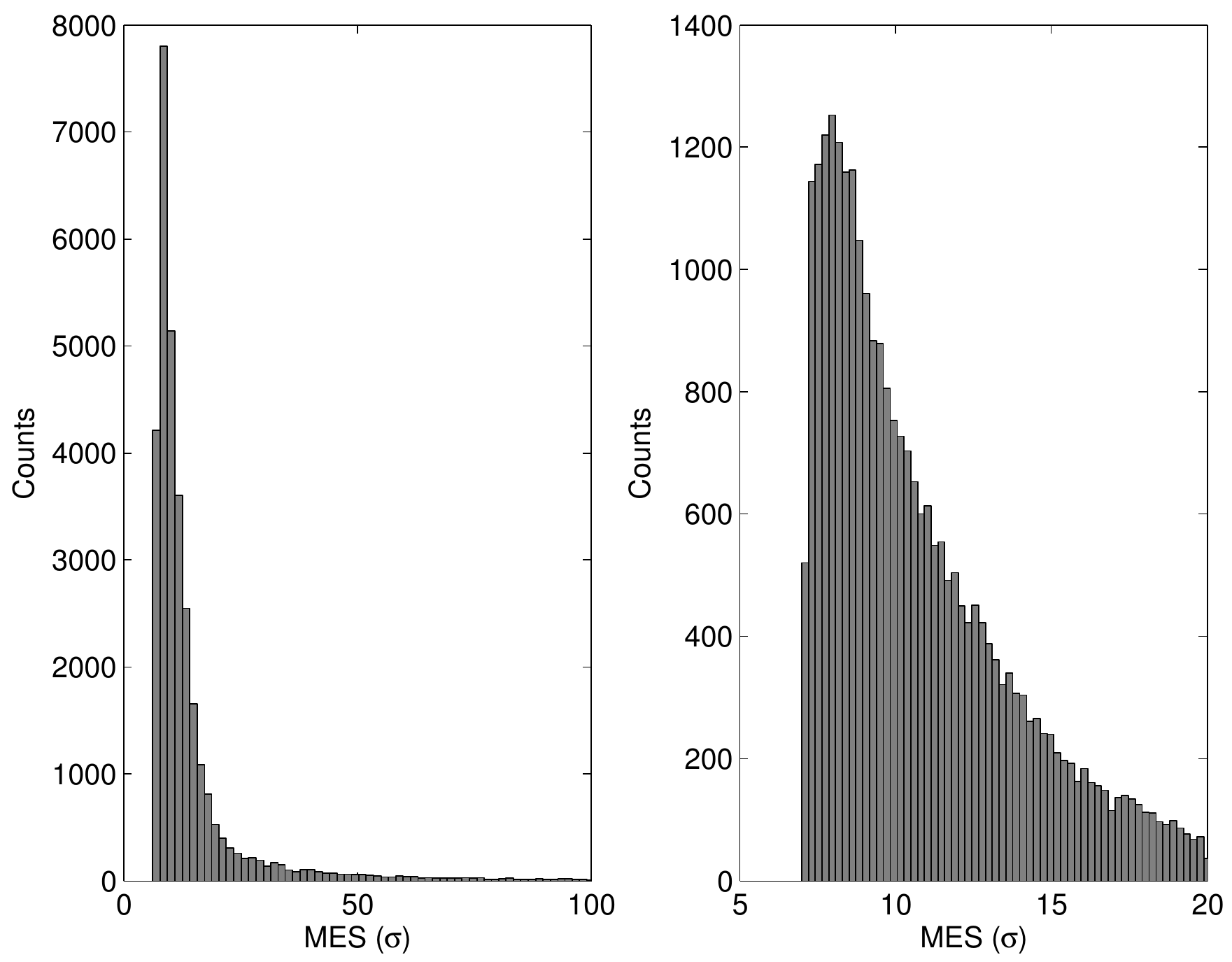}
\caption{Distribution of multiple-event statistics for all DR25 TCEs. Left: TCEs with MES below 100$\sigma$. Right: TCEs with MES below 20$\sigma$.
\label{f7}}
\end{figure}

Figure~\ref{f7} shows the distribution of DR25 MESs displayed on a linear scale. The high end of the MES distribution is clipped. In the left panel we see 31,064 TCEs with MES below 100$\sigma$; in the right panel we see 27,251 TCEs with MES below 20$\sigma$. The mode of the distribution is at 8$\sigma$, whereas \citet{seader2015} reported a mode near 9$\sigma$ in the Q1--Q17 DR24 results. Modifications to the TPS search algorithm and consistency tests have produced a population of TCEs for which the mode is 1$\sigma$ closer to the transit detection threshold.

\begin{figure}
%%\epsscale{.80}
%\plotone{fig8.eps}
\plotone{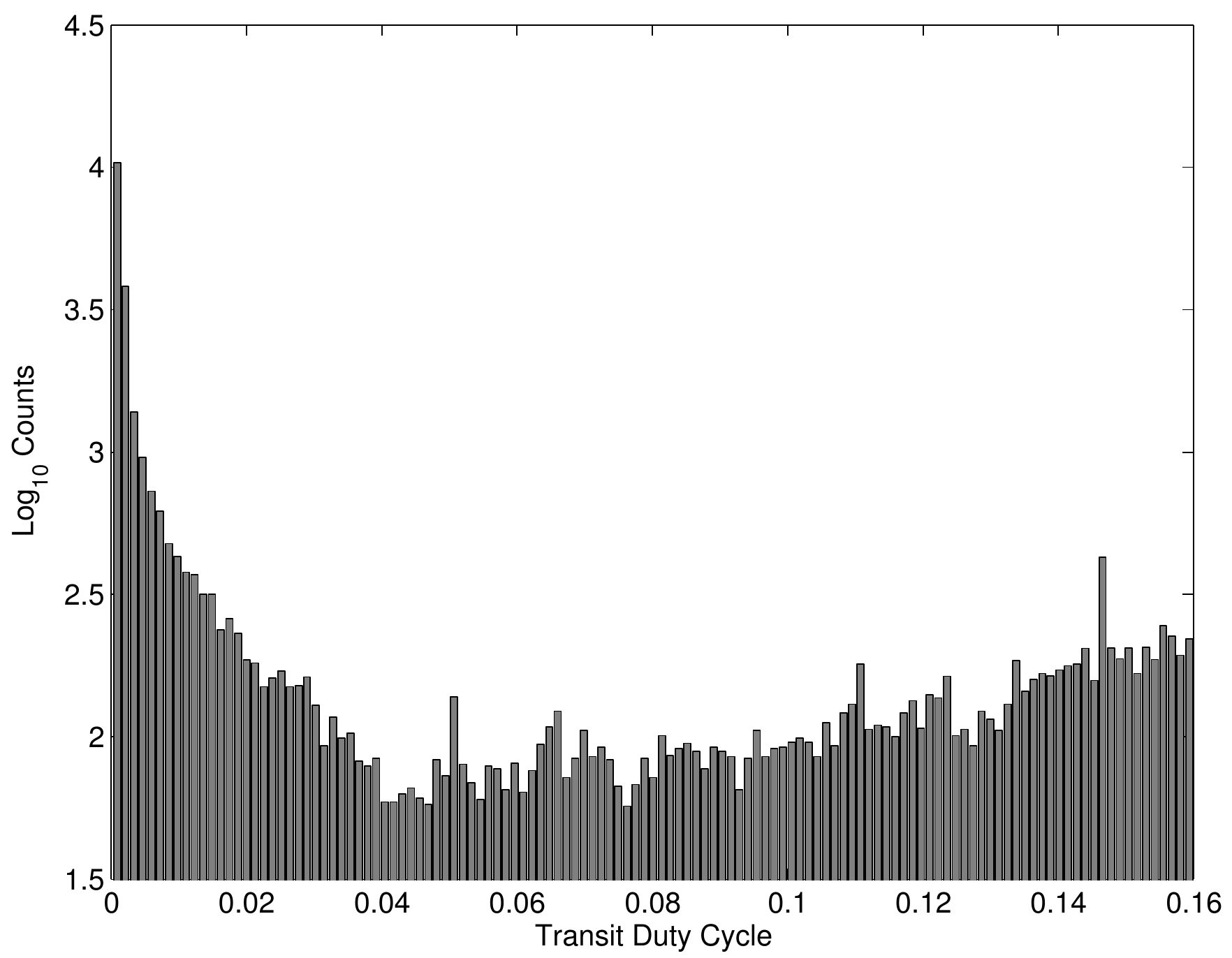}
\caption{Distribution of transit duty cycles on a logarithmic scale for the 34,032 DR25 TCEs. A duty cycle is defined as the ratio of the trial transit pulse duration to the detected period of the TCE.
\label{f8}}
\end{figure}

Figure~\ref{f8} shows a histogram of transit duty cycles for the DR25 TCEs.  The transit duty cycle is defined to be the ratio of the trial transit pulse duration to the detected period of the TCE (effectively the fraction of time during which the TCE is in transit). TPS employs a configurable upper limit on the duty cycles searched. The limit for this search was 0.16. This implies that the search for short-period transit signals does not include all of the trial pulse durations discussed earlier. The large number of short-period TCEs produces a ramp in duty cycle from 0.05 to 0.16, as described by \citet{seader2015}. The large number of long-period TCEs in the DR25 results, however, corresponds to many more TCEs at the lowest duty cycles than were generated in the DR24 run.

\subsection{Comparison with Known KOIs}
\label{ss:koi-comparison}
The performance of the latest Q1--Q17 pipeline run may be evaluated based on the rate of recovery of a set of well-established, high-quality KOIs. As in the past, we refer to these as ``golden KOIs.'' The ephemerides and dispositions for the golden KOIs were obtained from the cumulative KOI table at the NASA Exoplanet Archive on 2015 September 25, i.e., after the DR24 KOI table was finalized. The cumulative KOI table has been aggregated from past transit searches and \kepler{} planet catalogs published by \citet{borucki2,borucki3}, \citet{batalha1}, \citet{burke1}, \citet{rowe2015}, \citet{mullally2015}, and \citet{coughlin2016}. Selection criteria for the golden KOIs were (1)~MES above 9.0$\sigma$ in the most recent \kepler{} Pipeline transit search in which the KOI was recovered, and (2)~classification as PC following two or more prior transit searches including at least one of Q1--Q16 and Q1--Q17 DR24 . The DR25 golden KOI set includes 3402 KOIs on 2621 unique target stars. The size of this set far exceeds the number of golden KOIs employed in the past for evaluating pipeline transiting planet search performance. There were 1752 golden KOIs on 1483 target stars in the Q1--Q17 DR24 analysis, so the number of test cases for evaluating transit search performance nearly doubled for the final Q1--Q17 run.

TPS does not receive prior knowledge regarding KOIs or detections on specific targets. Recovery of objects of interest that were previously detected is a valuable test to guard against inadvertent introduction of significant flaws into the detection algorithm. The DR25 golden KOI set was selected for evaluation of the SOC 9.3 code base and associated pipeline runs approximately six months prior to the TPS and DV runs that produced the TCE population discussed here. Of the 3402 golden KOIs, 3385 (99.5\%) were classified in the cumulative KOI table as PCs at the time of the DR25 transit search. These include a number of PCs in systems featuring transit timing variations (TTVs). The pipeline was not designed to detect transit signatures with TTVs and has never been upgraded specifically for that purpose. Detection of transiting planets with TTVs represents a test of the robustness of the search algorithm, which assumes strictly periodic transit signals.

Whereas TPS does not receive prior knowledge regarding KOIs, DV is provided with the ephemerides (period, epoch, and transit duration) of the individual KOIs associated with each of the targets that produce TCEs in the transit search. KOI ephemerides are only employed by DV to match pipeline results for individual TCEs to known KOIs as an aid to \kepler{} project personnel and the greater science community. Matches to known KOIs (at the time that DV is run) are displayed in DV Reports by target and DV Report Summaries by TCE; these pipeline products are delivered to the NASA Exoplanet Archive. KOI ephemerides are not employed by DV for any purpose other than matching TCEs produced in the current pipeline run to previously known KOIs. An alternative matching algorithm is utilized outside of the pipeline to federate new TCEs with known KOIs when creating KOI tables at the Exoplanet Archive. It is possible (and even likely) that there will be discrepancies between KOI matching results displayed in the DR25 DV Reports and DV Report Summaries and the DR25 KOI table at the Archive. The matching of TCEs to KOIs is discussed further in Section~\ref{ss:matching}.

It should be noted that PDC does employ ephemerides of known KOIs and eclipsing binaries in conditioning data for the transiting planet search. The ephemerides are used to identify in-transit (or in-eclipse) cadences for given targets; data samples for those targets and cadences are subsequently protected from misidentification as outliers or sudden pixel sensitivity dropouts \citep{jenkins2010a,stumpe2012}.

\begin{figure}
%%\epsscale{.80}
%\plotone{fig9.eps}
\plotone{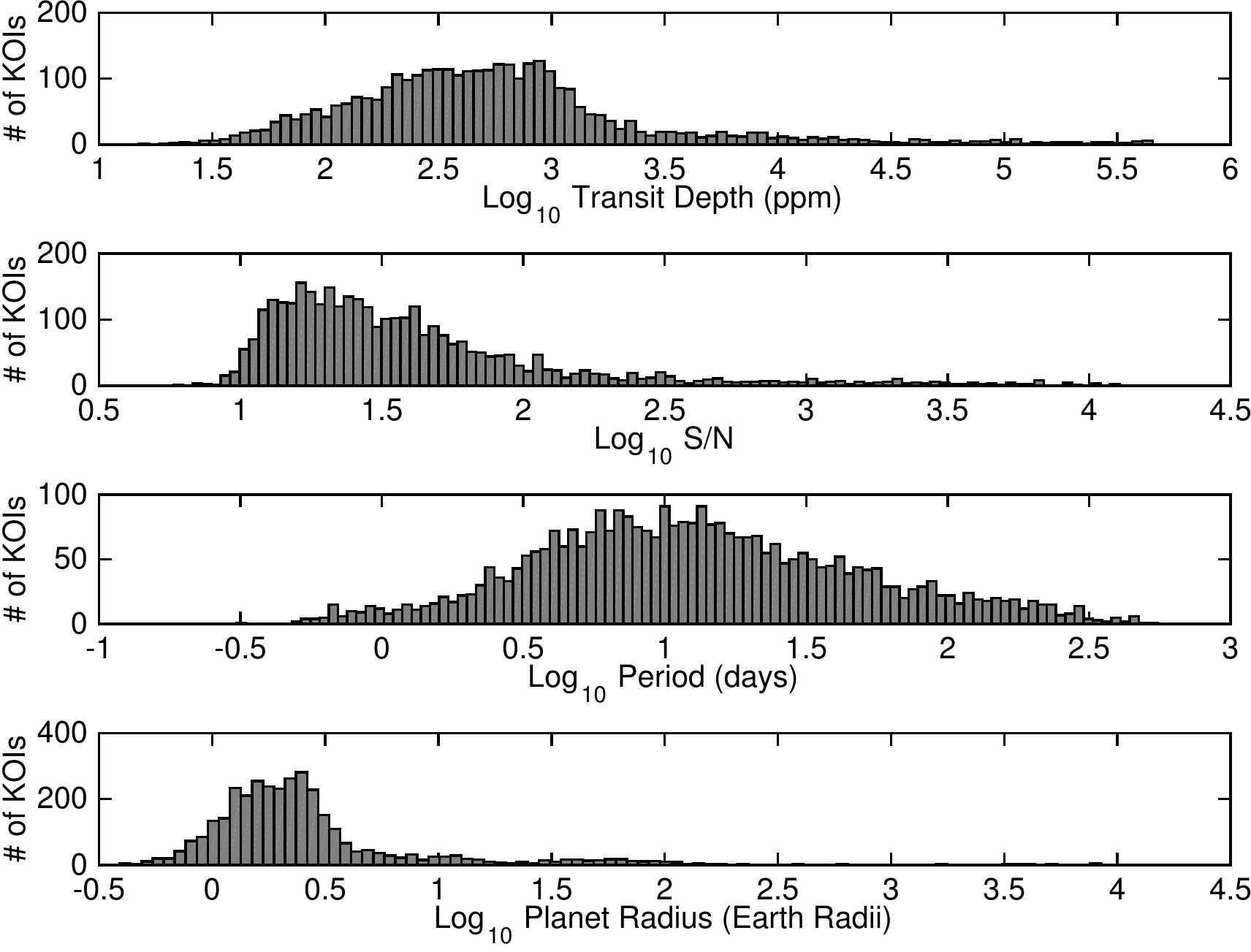}
\caption{Parameter distributions of ``golden KOIs''  employed to assess performance of DR25 transit search. Parameters are displayed on logarithmic horizontal axes. Parameter values were obtained from the cumulative KOI table at the NASA Exoplanet Archive on 2015 September 25.
\label{f9}}
\end{figure}

Figure~\ref{f9} shows the distribution of transit depth, signal-to-noise ratio, period, and planet radius for the DR25 golden KOIs. Parameters are displayed on logarithmic scales. As discussed earlier, the parameter values were obtained from the cumulative KOI table at the NASA Exoplanet Archive on 2015 September 25. The golden KOIs span a large region of transiting planet parameter space. In the final pipeline run, TPS produced TCEs on targets hosting 3397 of the 3402 golden KOIs. TCEs were not produced on targets hosting the following golden KOIs:
\begin{enumerate}
\item 4253.01 (period 173.3 days, S/N 13.5)
\item 4670.01 (period 6.81 days, S/N 11.5)
\item 4886.01 (period 18.0 days, S/N 12.1)
\item 5727.01 (period 65.4 days, S/N 9.6)
\item 5850.01 (period 303.2 days, S/N 12.7)
\end{enumerate}

The TPS search latched onto the correct period for all five of these KOIs. The reasons for failure to produce TCEs on these targets are as follows:
\begin{enumerate}
\item Maximum MES detection statistic (7.04$\sigma$) was below 7.1$\sigma$ threshold (KOI 5727.01).
\item Veto by $\chi^2_{(GOF)}$ consistency test (4670.01, 4886.01, and 5850.01).
\item Veto by both $\chi^2_{(2)}$ and $\chi^2_{(GOF)}$ consistency tests (4253.01).
\end{enumerate}

The $\chi^{2}$ vetoes were discussed earlier and described in detail by \citet{seader2013}. We will now discuss matching of KOI and TCE ephemerides for the targets hosting golden KOIs on which TCEs were generated in the DR25 run.

\subsection{Matching of Golden KOI and TCE Ephemerides}
\label{ss:matching}
Generation of a TCE on a target hosting a golden KOI is not sufficient to determine that the KOI has been recovered. It is necessary that the transit ephemeris produced in the pipeline match that of the KOI, that is, that the transit events producing the pipeline detection are consistent with the KOI ephemeris. As stated earlier, DV is provided with the ephemerides (period, epoch, transit duration) of each of the previously known KOIs associated with the targets that produce TCEs in the search for transiting planets. Once the limb-darkened model fitting and search for additional planets have concluded, DV matches each of the pipeline results against the ephemerides of the known KOIs. The matching is performed by correlating high temporal resolution, rectangular KOI transit waveforms against rectangular transit waveforms based on pipeline ephemerides;  Pearson correlation coefficients are  compared against an ephemeris matching threshold (0.75 for KOI matching in DV) to establish whether or not a match has been produced. The transit waveforms are normalized such that correlation coefficients equal to 1.0 indicate perfect matches and correlation coefficients equal to 0.0 indicate that KOI and pipeline transit events are nonoverlapping. The matching threshold was specified at a level that is likely to be reached only if a KOI is accurately recovered. A match is not declared in DV if the correlations exceed the matching threshold between a single TCE and multiple KOIs on a given target (which does happen in the case of duplicate KOIs), or if the correlations exceed the matching threshold between multiple TCEs on a given target and a single KOI.

In the DR25 run, DV reported an ephemeris match at the specified threshold or better for 3354 of 3402 golden KOIs. These golden KOIs may be assumed to have been recovered without further investigation. Of the 48 golden KOIs that did not trigger an ephemeris match, we have seen that in five cases there was no TCE on the host target. We investigated the remaining 43 golden KOIs for which a TCE was produced on the host target but a match was not reported by DV in order to ascertain whether the KOIs in question had, in fact, been recovered. We found that 40 of the 43 golden KOIs in this category were indeed recovered in the DR25 run; details are provided below. In total, 3394 of 3402 DR25 golden KOIs (99.8\%) were recovered in the pipeline with the SOC 9.3 code base.

\begin{figure}
%%\epsscale{.80}
%\plotone{fig10.eps}
\plotone{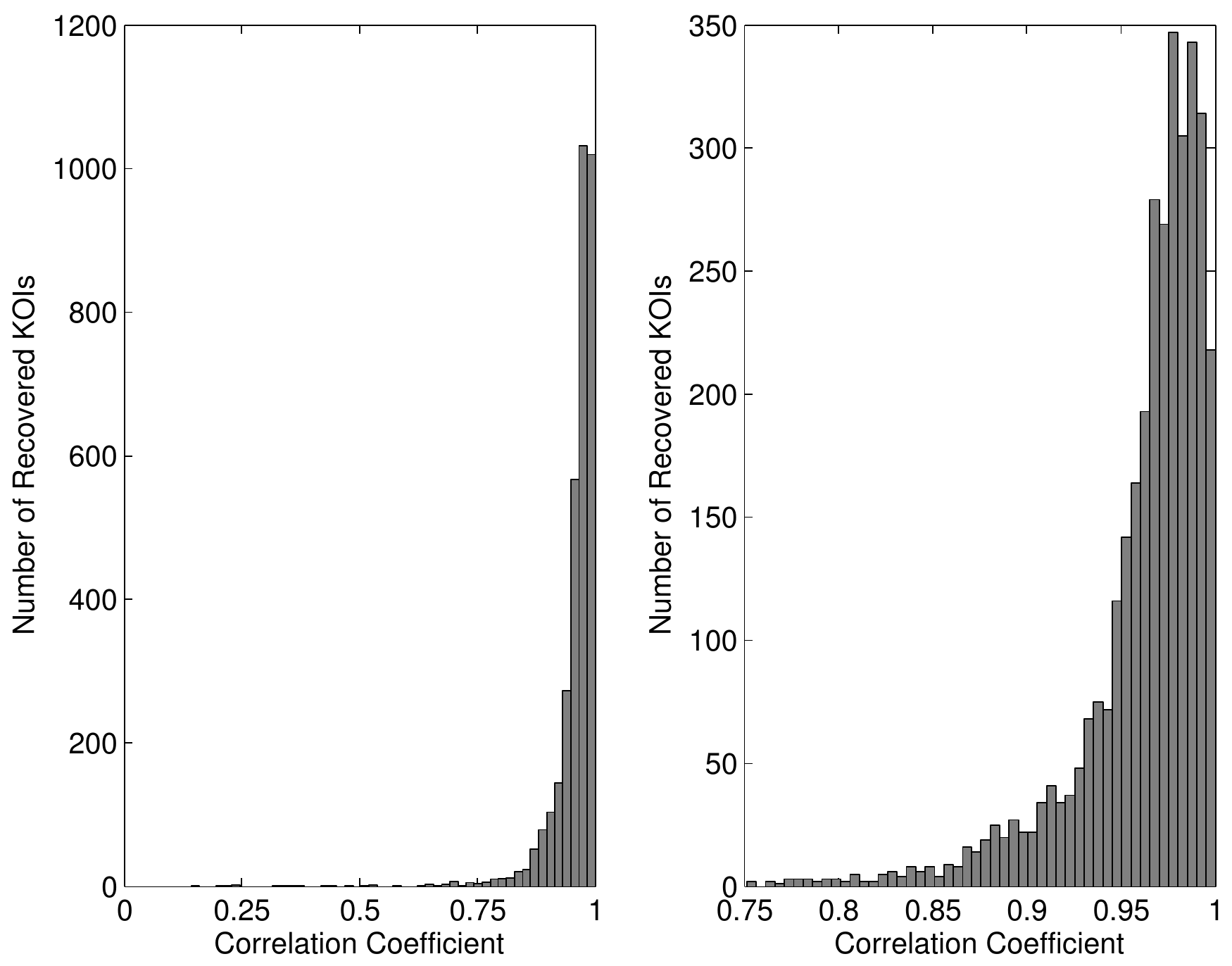}
\caption{Distribution of ephemeris match correlation coefficients for ``golden KOIs'' recovered in DR25 run. Left: all recovered KOIs. Right: recovered KOIs at ephemeris matching threshold (0.75) and above.
\label{f10}}
\end{figure}

Figure~\ref{f10} shows the distribution of ephemeris match correlation coefficients for the 3394 golden KOIs that were recovered in the DR25 pipeline run. The full range of correlation coefficients is displayed in the left panel, and the range of correlation coefficients above the pipeline ephemeris matching threshold (0.75) is displayed in the right-hand panel. Most of the ephemerides were matched at a high level. In fact, 92.0\% of the golden KOIs that were recovered in the run produced correlation coefficients $> 0.9$.

Table~\ref{t3} lists the complete golden KOI set, KOI and TCE ephemerides, and corresponding ephemeris match correlation coefficients; the table is sorted by KOI number. The time standard for the epoch of first transit is the barycentric \textit{Kepler}-modified Julian date (BKJD). It should be noted that DV employs a standard convention for reporting epochs (first transit after start of Q1), whereas the cumulative KOI table at the Exoplanet Archive does not; hence, there are often an integer number of orbital periods between the KOI and TCE epochs. TCE ephemerides and correlation coefficients are listed as -1 for the eight golden KOIs that were not recovered in the DR25 run.

The five golden KOIs for which no TCE was produced on the host target were discussed earlier. The three unrecovered golden KOIs for which at least one TCE was produced on the host target are as follows:
\begin{enumerate}
\item 989.03 (period 16.2 days, S/N 52.8)
\item 2048.02 (period 99.7 days, S/N 12.6)
\item 3051.01 (period 11.7 days, S/N 11.2)
\end{enumerate}
The reasons for failure to produce TCEs for these KOIs follow:
\begin{enumerate}
\item Self-timeout in DV model fitting process on earlier TCE associated with FP KOI on target (KOI 989.03)
\item Veto by both $\chi^2_{(2)}$ and $\chi^2_{(GOF)}$ consistency tests after latching onto correct period (2048.02 and 3051.01)
\end{enumerate}

We explain below why a match was not reported at the specified threshold in DV for many of the 40 golden KOIs that were recovered nonetheless to provide a sense of the difficulties involved in benchmarking pipeline completeness by matching TCE to KOI ephemerides:
\begin{enumerate}
\item KOI and DV periods differ by an integer factor and the KOI period appears to be incorrect (2174.03, 4829.01, 4893.01).
\item KOI and DV periods differ by an integer factor and the DV period appears to be incorrect (2306.01, 2732.04).
\item KOI and DV periods differ by an integer factor but the true period is ambiguous because the target was observed only in every other quarter (5568.01).
\item KOI period is less than 0.5~days and DV produces two TCEs at twice the true period because the minimum search period is 0.5~days (2916.01).
\item KOI transit duration is less than one cadence; the minimum TPS trial transit duration is 1.5~hr (three cadences), and the DV fitter declares an error if the duration derived from fit parameters is less than one cadence (4546.01).
\item KOI is in a TTV system \citep{ford2012,mazeh2013} where linear KOI and pipeline ephemerides only approximate the observed transit signal (KOI 277.02, 456.02, 884.02, 984.01, 1831.03).
\item KOI was observed late in mission only; a discrepancy exists between KOI and DV epochs when projected back to the start of science operations (5403.01, 5605.01, 5672.01, 6145.02, 6166.02).
\item Duplicate KOIs exist; DV does not report a match because the pipeline result actually matches two different KOIs on the same target (1101.01, 2768.01).
\item KOI is binary \citep{eb-cat3}, featuring deep eclipses ($> 25$\%) that DV does not fit by design; the TPS ephemeris does not match KOI at the specified threshold (3545.01, 3554.01, 5797.01).
\item KOI appears to be heartbeat star \citep{eb-cat3} that does not feature conventional transits or eclipses, but rather rings due to tidal pulsations (2215.01).
\end{enumerate}

In summary, 3394 of 3402 golden KOIs (99.8\%) selected in advance for assessment of the performance of the DR25 run with the SOC 9.3 code base were recovered in the pipeline. Of the eight golden KOIs that were not recovered, in one case the correct period was latched but there was a failure to produce a detection statistic at the required threshold (7.04 versus 7.1$\sigma$), in six cases the correct period was latched and a sufficient detection statistic was produced but a TCE was vetoed by one or two of the $\chi^2$ transit consistency tests, and in one case a timeout limit was reached while processing a prior FP TCE in a multiple-KOI system.

The vetoes are a necessary evil for the transiting planet search. They impact completeness to some degree. Without the vetoes, however, the sheer number of TCEs would overwhelm the ability to process them in the pipeline and to produce a reliable catalog of planetary candidates. As stated earlier, the DR25 run would have produced TCEs on 68,004 unique targets at the 7.1$\sigma$ level or greater in the absence of the vetoes. This would have approximately doubled to 136,000 total TCEs after the search for additional planets. A larger number of TCEs does not necessarily imply better completeness, however; the vetoes allow a continued search of light curves without passage to DV and removal of flux data in the pipeline search for multiple planets on individual targets. It is interesting to note that there were 126,153 unique targets with MES above 7.1$\sigma$ prior to application of the vetoes in the DR24 run \citep{seader2015}; this is significantly larger than the 68,004 targets with MES above 7.1$\sigma$ in the DR25 run. We believe that the reduction in DR25 is likely attributable to implementation of quarter-by-quarter whitening in TPS, as discussed earlier.

The 99.8\% recovery rate with the SOC 9.3 code base represents exceptional performance for a large set of KOIs. \citet{seader2015} reported that 1664 of 1752 golden KOIs (95.0\%) were recovered in the Q1--Q17 DR24 run with the SOC 9.2 code base. \citet{pt2014} reported that 1597 of 1646 golden KOIs (97.0\%) were recovered in the Q1--Q16 run with the SOC 9.1 code base.

Although recovery of a large set of established KOIs is a solid test of the performance of the pipeline search for transiting planets, it should be noted that KOIs do not reflect ground truth because the true nature of these objects of interest is not actually known. A better measure of the detection efficiency of the pipeline is the recovery of transiting planet signatures injected into \kepler{} flight data for a large number of targets over a range of orbital periods and planet radii (and hence S/N). Such studies have been performed for  four quarters (Q9--Q12) with the SOC 9.1 code base \citep{christiansen2015} and 17~quarters with the SOC 9.2 code base \citep{christiansen2016a}. These studies involve the injection of transit signatures into calibrated pixel data and subsequently running the \kepler{} Pipeline through the PA, PDC, TPS, and DV components. Pixel-level injections also offer the opportunity to characterize the performance of the DV diagnostics employed to help differentiate between true transiting planet signatures and FP detections attributable to eclipsing binaries and background sources \citep{bryson2013,mullally2015,coughlin2016}. A pixel-level transit injection run is currently underway to assess the detection efficiency of the DR25 pipeline \citep{christiansen2016b}.

Parallel studies are also underway involving the injection of transit signals into the systematic error-corrected light curves of a representative sample of long-cadence targets. The flux-level injections are repeated many times (typically $> 600,000$) for each target in order to deeply probe planet search detection efficiency over a range of orbital periods and planet radii. Both pixel-level and flux-level injections represent a better gauge of pipeline completeness than recovery of existing KOIs because these are controlled experiments where a very large number of transit signatures may be injected and ground truth is available. Nevertheless, transit injections require considerable time and computational resources; assessing pipeline performance based on recovery of established KOIs remains a valuable exercise and is the best way to characterize performance quickly from one transit search to the next.

\section{DV RESULTS}
\label{s:dv}
The DV module characterizes the planet and orbital parameters associated with each TCE, removes modeled transit events and identifies transit signatures in target light curves beyond those initially found in TPS, and performs a suite of diagnostic tests to aid in discrimination between true transiting planets and false alarm/false positive detections. The DV module and overall TCE validation approach are discussed in \citet{hw2010} and \citet{jdt2015,jdt2016}. The DV transiting planet model fitting algorithm and multiple planet search are described in \citet{pt2010} and \citet{li2015,li2016}.

There were substantial upgrades to DV functionality in the final revision of the pipeline code base (SOC 9.3). We list them here but defer the details to \citet{jdt2016} and \citet{li2016}:
\begin{enumerate}
\item Trapezoidal model fitting to provide fast, robust model fits
\item Fallback to trapezoidal model to support diagnostic tests when limb-darkened model fit result is not available (6.7\% of total DR25 TCEs)
\item Computation of effective stellar flux (i.e.,~insolation relative to solar flux received at the top of Earth's atmosphere) at a distance specified by the semimajor axis
\item Propagation of uncertainties in stellar parameters to uncertainties in fitted and derived parameters
\item Fitted depth and associated uncertainty for the strongest secondary event at the period and pulse duration associated with the TCE
\item Computation of geometric albedo and comparison statistic against 1.0 to distinguish between planetary occultations and secondary eclipses. The geometric albedo $A_{g}$ is computed by
\begin{equation}
%A_{g} = D \: \bigg({a_{p} \over R_{p}}\bigg)^{2},
A_{g} = D \: \bigg(\frac{a_{p}}{R_{p}}\bigg)^{2},
\end{equation}
where $D$ is the fractional depth of the strongest secondary event at the period and pulse duration associated with the TCE, $a_{p}$ is the semimajor axis of the orbit, and $R_{p}$ is the planet radius.
\item Computation of planet effective temperature and comparison statistic against planet equilibrium temperature to distinguish between planetary occultations and secondary eclipses. We estimate the effective temperature of the orbiting companion by assuming the flux from the companion is thermal (blackbody) emission and that the measured occultation depth is approximately equal to the ratio of the planet to star luminosity. The planet effective temperature $T_{p}$ is then determined by
\begin{equation}
T_{p} = T_{*} \: D^{1/4} \: (R_{p} / R_{*})^{-1/2},
\end{equation}
where $T_{*}$ is the effective temperature of the host star, $D$ is the fractional depth of the strongest secondary event at the period and pulse duration associated with the TCE, and $(R_{p} / R_{*})$ is the fitted reduced-radius parameter.
\item Rolling band contamination diagnostic to identify coincidences between transit events and rolling band artifacts
\item Optical ghost diagnostic test to identify cases where detections are triggered by optical ghosts (or other means of distributed contamination)
\item Two-dimensional statistical bootstrap diagnostic to assess the false-alarm probability associated with each detection
\item Quarterly, annually, and seasonally phased light curves in DV Report
\item Redesigned DV Time Series data product for delivery to Exoplanet Archive \citep{thompson2016}
\end{enumerate}

This functionality applies to all TCEs identified in the DR25 run. The new features enhance the quality of the diagnostics produced by DV to support TCE vetting and the presentation of DV results in reports and summaries employed in the vetting process and delivered to the Exoplanet Archive. We will now discuss quality and reliability of the Q1--Q17 DR25 TCEs and then characteristics of the planet population implied by the transiting planet model fits.

\subsection{TCE Quality and Reliability}
\label{ss:quality}
A limb-darkened transiting planet model fit is employed in DV to search for the best astrophysical model across the parameter space of orbital period, epoch of first transit, impact parameter, ratio of planet radius to host star radius (reduced radius), and ratio of orbital semimajor axis to host star radius (reduced semimajor axis). The astrophysical models are constructed using the geometric transit model of Mandel and Agol \citep{mandel} with nonlinear Claret limb darkening \citep{claret2011}. The stellar parameters for the DR25 run were provided by the \kepler{} Stellar Properties Working Group \citep{huber2016}; the parameters employed in DV are stellar radius, effective temperature, surface gravity (log g), and metallicity (log Fe/H).

The S/N of the model fit should, in general, be slightly larger than the MES detection statistic that TPS returns, for a number of reasons: (1)~the ephemeris is more refined, (2)~the signal--template match is better (due to the use of barycentric corrected time stamps and target-specific astrophysical parameters), and (3)~whitener performance is improved because in-transit cadences are known in advance. A fit S/N that is significantly different from the MES may indicate that the TCE was produced by some phenomenon other than a planet transit or eclipse. A fit S/N that is below the pipeline detection threshold (7.1$\sigma$) calls the validity of the TCE into question.

\begin{figure}
%%\epsscale{.80}
%\plotone{fig11.pdf}
\includegraphics[bb=32 17 501 393]{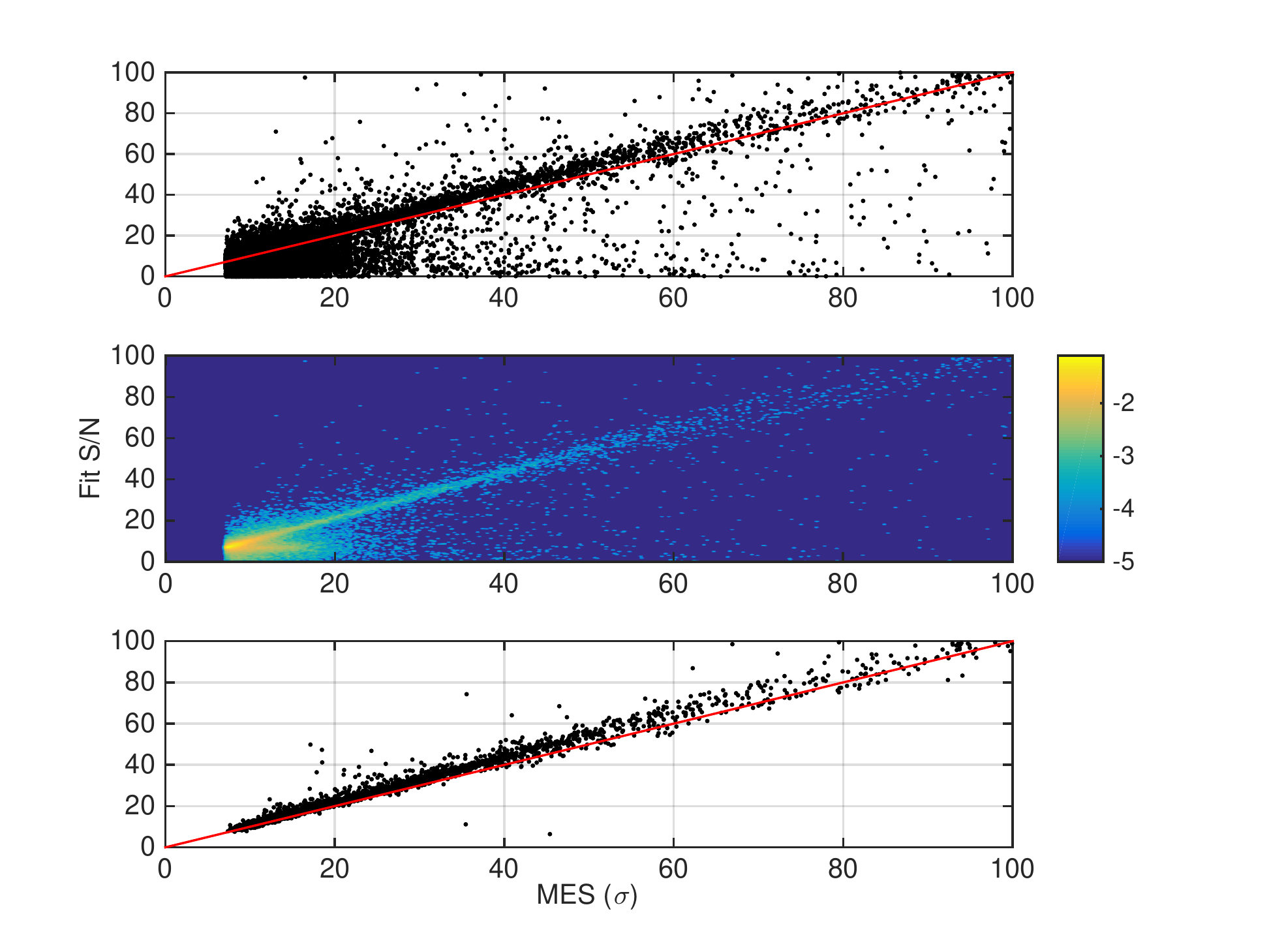}
\caption{DV model fit S/N vs. TCE multiple-event statistic. Red lines denote model fit S/N = MES. Top: all DR25 TCEs.  Middle: density for all TCEs on logarithmic scale. Bottom: ``golden KOIs.'' Figures display MES below 100$\sigma$.
\label{f11}}
\end{figure}

Figure~\ref{f11} compares the S/N for the limb-darkened transiting planet model fit and the MES detection statistic for all DR25 TCEs (with model fit results) in the top panel. The density of the S/N-versus-MES distribution is shown in the middle panel. The S/N for the model fit and the MES detection statistic for TCEs associated with golden KOIs are displayed in the bottom panel. The axes have been restricted to focus on the bulk of the population. The golden KOIs exhibit the behavior expected for a population consisting primarily of transiting PCs. The S/N is generally larger than the MES across the full range of detection statistics displayed in the figure. The model fit S/N is below the pipeline detection threshold (7.1$\sigma$) for one of the golden KOIs; this KOI (884.02, also known as Kepler-247d) features TTVs on the same order as the transit duration. The transit data are smeared when folded at the nominal period, and a low-S/N fit results. There are 7905 TCEs in the full DR25 population for which the model fit S/N is below the pipeline threshold. The reliability of these TCEs is called into question.

\begin{figure}
%%\epsscale{.80}
%\plotone{fig12.pdf}
\includegraphics[bb=21 17 510 393]{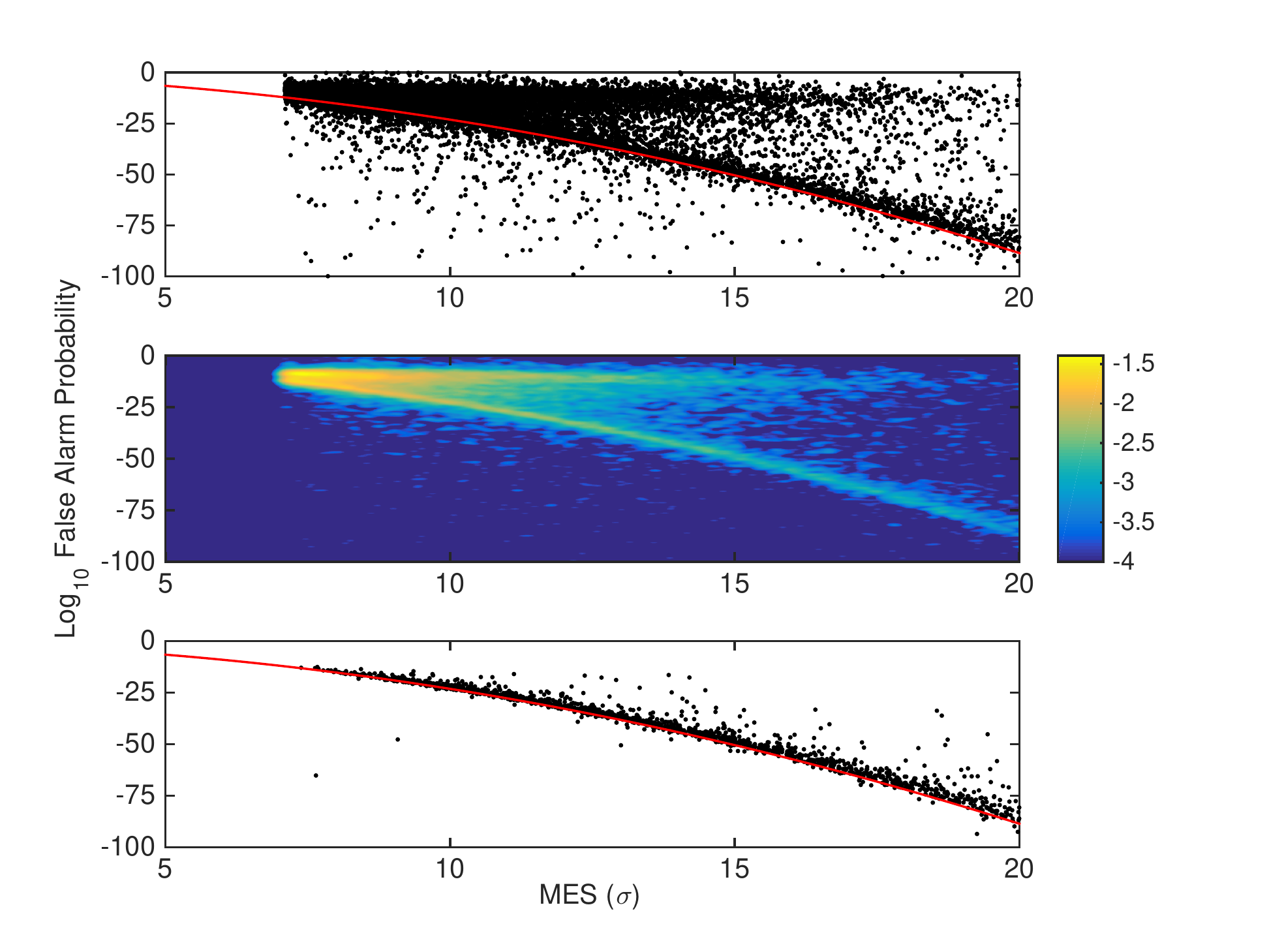}
\caption{DV bootstrap false alarm probability (on logarithmic scale) vs. TCE multiple-event statistic. Red curves denote false-alarm probability under assumption of Gaussian noise statistics. Top: all DR25 TCEs.  Middle: density for all TCEs on logarithmic scale.  Bottom: ``golden KOIs.'' Figures display MES below 20$\sigma$.
\label{f12}}
\end{figure}

Figure~\ref{f12} displays the false-alarm probability determined by the 2D statistical bootstrap \citep{jenkins2015} in DV against the TCE MES. The false-alarm probability represents the likelihood that a detection statistic as large or larger than the MES associated with the TCE would have been produced with the same number of transit events and the same trial transit duration in the absence of all transits (associated with all TCEs) on the target star. The false-alarm probabilities for the DR25 TCEs (for which the bootstrap was computed) are shown on a logarithmic scale in the upper panel in the figure. The density of the false-alarm probability versus MES distribution is shown in the middle panel. The false-alarm probabilities for the TCEs associated with golden KOIs are shown on a logarithmic scale in the lower panel. The axes have been restricted to focus on the bulk of the population. Red curves indicate the false-alarm probability under the assumption of Gaussian noise statistics.

The golden KOIs track the Gaussian expectation for false-alarm probability with MES (recall that the whitener employed in TPS is designed to produce standard normal noise statistics). The full TCE population exhibits a horizontal band representing a large number of TCEs with false-alarm probabilities well above the expectation for Gaussian noise statistics. These TCEs appear to be well separated from the bulk of the remaining TCE population, and their validity should be investigated. In the full TCE population, there are 8634 TCEs with false-alarm probabilities $> 10^{-12}$ (of which 7206 exceed $10^{-11}$ and 5498 exceed $10^{-10}$); in the golden KOI population there are none. The 7.1$\sigma$ transit search detection threshold was chosen to yield a false-alarm probability of $6.2 \rm{x} 10^{-13}$ for a properly whitened noise distribution. The number of independent statistical tests in the transiting planet search for roughly 200,000 target stars with four years of observations is on the order of $10^{12}$ \citep{jenkins2002a}.

\begin{figure}
%%\epsscale{.80}
%\plotone{fig13.pdf}
\includegraphics[bb=28 17 510 393]{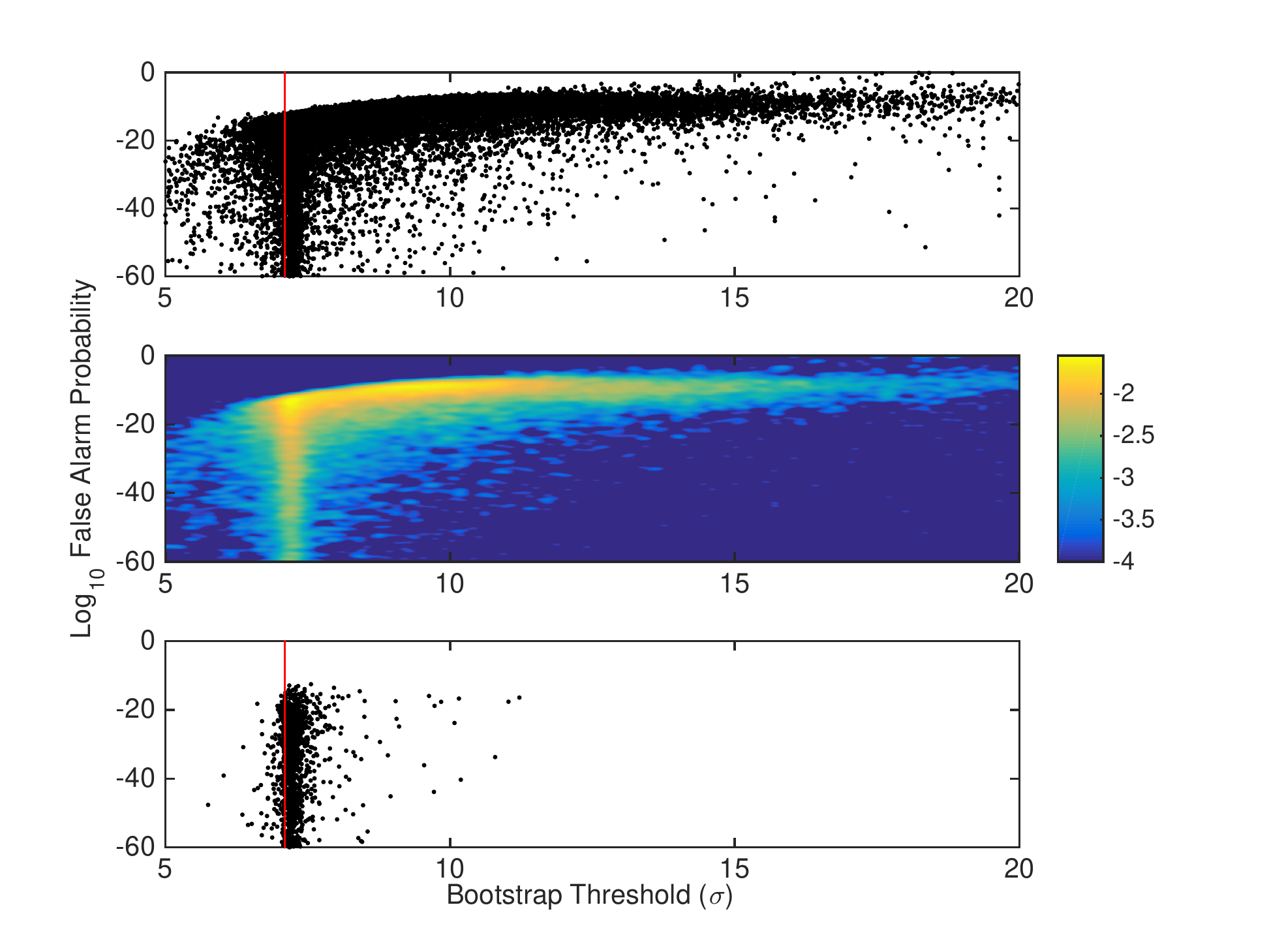}
\caption{DV bootstrap false-alarm probability (on logarithmic scale) vs. bootstrap threshold. Red lines denote threshold = 7.1$\sigma$. Top: all DR25 TCEs.  Middle: density for all TCEs on logarithmic scale.  Bottom: ``golden KOIs.'' Figures display bootstrap threshold below 20$\sigma$ and false-alarm probability above $10^{-60}$.
\label{f13}}
\end{figure}

Figure~\ref{f13} displays the false-alarm probability against the statistical bootstrap threshold. The bootstrap threshold represents the transit detection threshold that would have been required to produce the false-alarm probability equivalent to a 7.1$\sigma$ threshold on a Gaussian noise distribution. The false-alarm probabilities for the DR25 TCEs are shown on a logarithmic scale in the upper panel in the figure. The density of the false-alarm probability versus bootstrap threshold distribution is shown in the middle panel. The false-alarm probabilities for the TCEs associated with golden KOIs are shown on a logarithmic scale in the lower panel. The axes have been restricted to focus on the bulk of the population. The golden KOI population exhibits the desired behavior; the bootstrap threshold is essentially constant near 7.1$\sigma$ (shown in red) over the range of displayed false-alarm probabilities. A large population of false alarms is evident in the top panel, however, for which the noise statistics are non-Gaussian, and a higher detection threshold would have been required to maintain a uniform false-alarm rate.

\subsection{Planet Characteristics}
\label{ss:characteristics}
Transiting planet signatures for small planets (relative to the size of their host stars) may be fitted well over a wide range of impact parameters. In recent DV revisions, we have attempted to uniformly distribute the initialization seed for the fitted impact parameter to some degree. The method for seeding the iterative, limb-darkened transiting planet model fits was modified in the SOC 9.3 code base as described by \citet{li2016}. An inadvertent side effect of the change in model fit seeding is that all fits were initialized with a high impact parameter (0.9). This did not bias the DR25 results relative to DR24 for TCEs corresponding to large planets, but it did somewhat bias the DR25 results relative to DR24 for TCEs corresponding to small planets. Convergence for large planets was essentially independent of impact parameter seed, whereas such was not the case for small planets. Transiting planets with a high impact parameter must be larger than those with lower impact parameter for given transit depths on the same host stars because of limb darkening.

\begin{figure}
%%\epsscale{.80}
%\plotone{fig14.eps}
\plotone{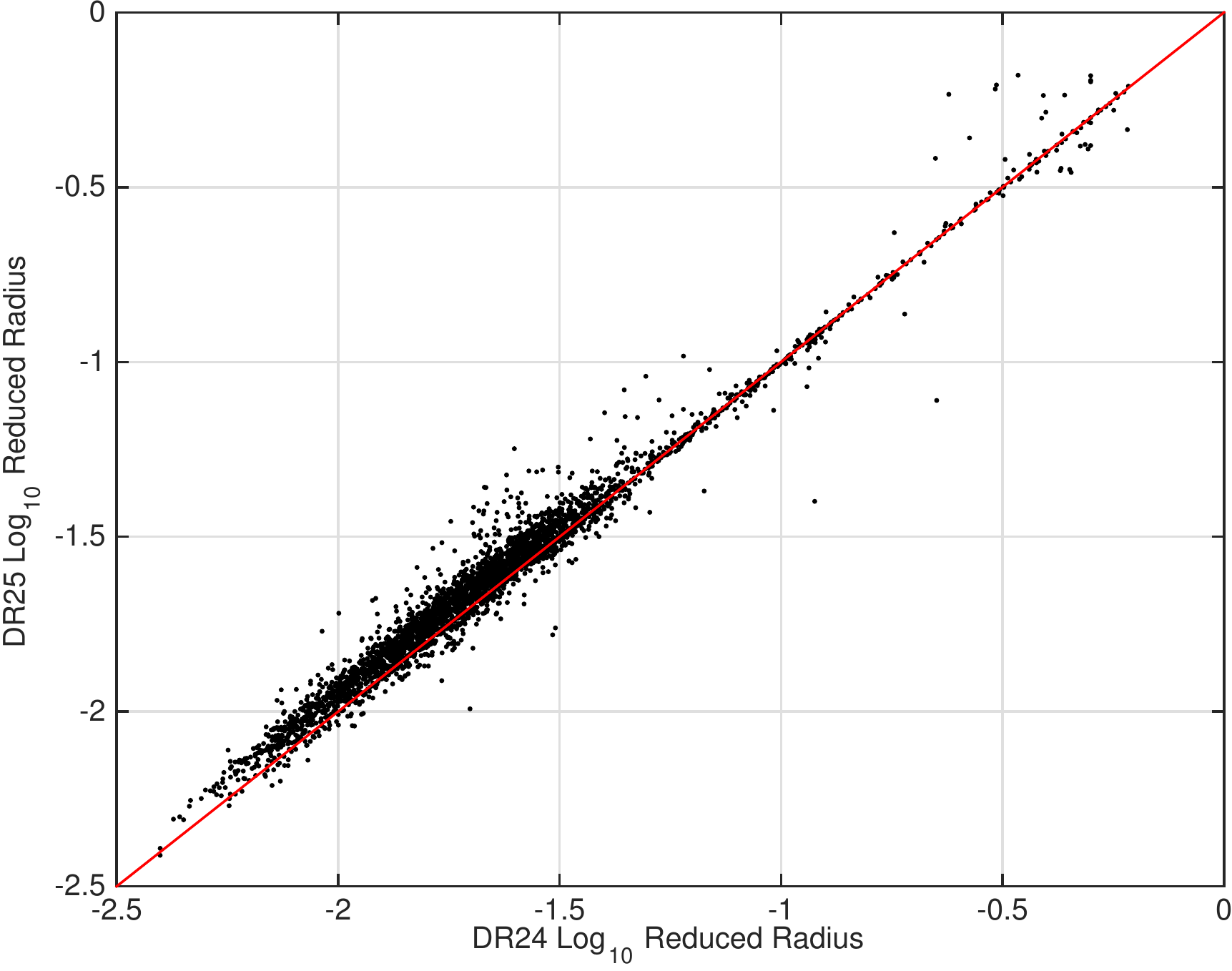}
\caption{DR25 reduced-radius ($R_p/R_*$) fit parameter vs. DR24 reduced-radius fit parameter on logarithmic scales for ``golden KOIs'' recovered in both pipeline runs. The red line denotes DR25 reduced radius = DR24 reduced radius. There is no bias between the two sets of results for TCEs with reduced radius $> 0.05$ (log\textsubscript{10} = -1.3). The median increase from DR24 to DR25 in reduced radius for TCEs with (DR24) reduced radius $< 0.05$ is 9.8\%.
\label{f14}}
\end{figure}

Figure~\ref{f14} displays the DR25 reduced-radius ($R_p/R_*$) fit parameter against the DR24 reduced radius fit parameter for golden KOIs that were recovered in both pipeline runs. The reduced radius parameters are displayed on logarithmic scales. There is no bias between the two sets of results for TCEs with reduced radius $> 0.05$ (log\textsubscript{10} = -1.3). The median increase in reduced radius from DR24 to DR25 for TCEs with reduced radius $< 0.05$ is 9.8\%. By contrast, the median fractional uncertainty in derived planet radius for the same DR25 golden KOI TCEs is 29.8\%. All planetary candidates in the DR25 \keplermission{} catalog will be modeled independently by TCERT, so the bias discussed here relates specifically to TCE products at the Exoplanet Archive.

\begin{figure}
%%\epsscale{.80}
%\plotone{fig15.pdf}
\includegraphics[bb=34 14 501 393]{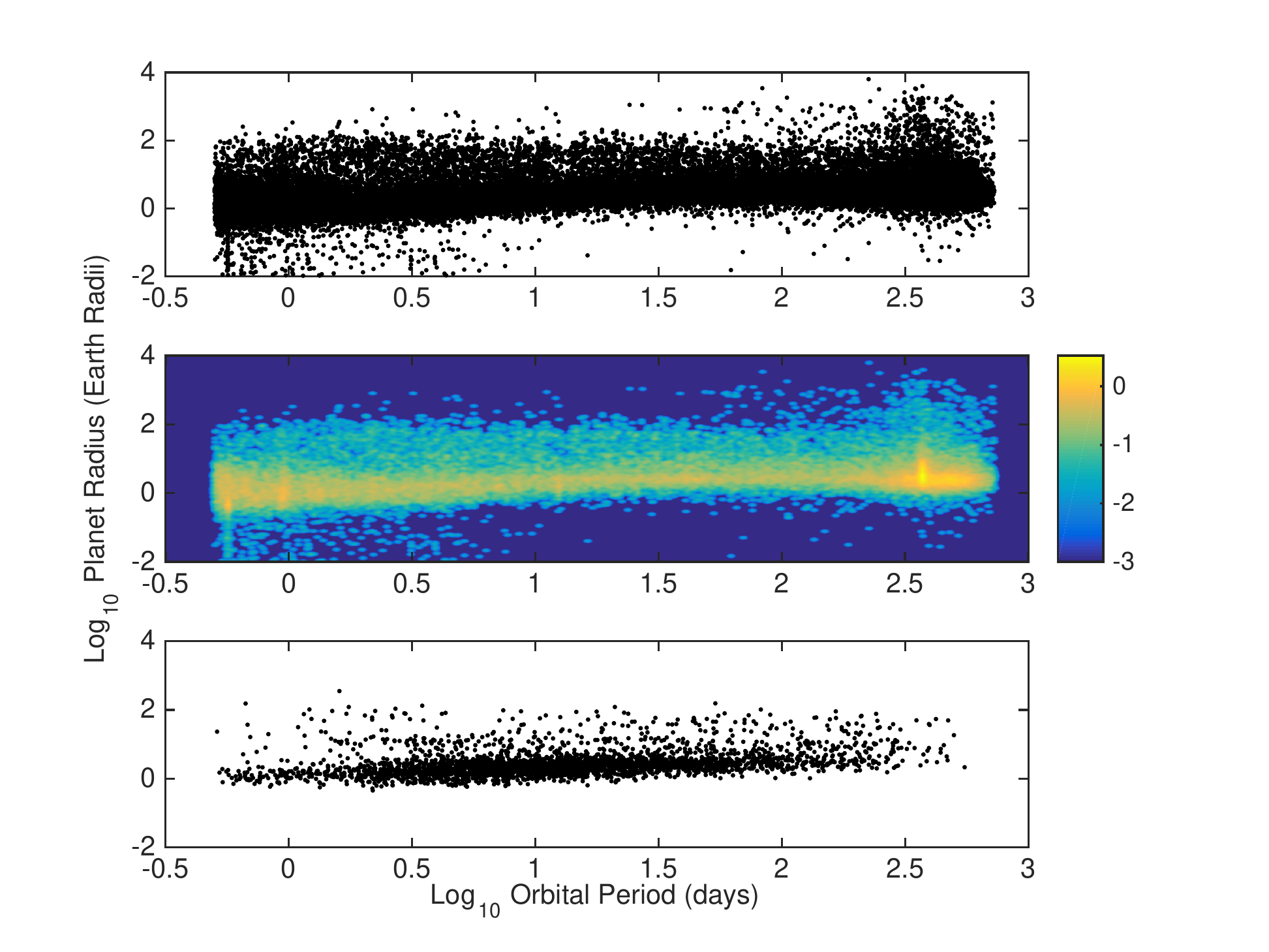}
\caption{Planet radius in \rearth{} vs. orbital period in days on logarithmic scales. Top: all DR25 TCEs. Middle: density for all TCEs on logarithmic scale. Bottom: ``golden KOIs.''
\label{f15}}
\end{figure}

The planet radius $R_{p}$ associated with each TCE is derived in DV from the limb-darkened transiting planet model fit result by
\begin{equation}
R_{p} = (R_{p} / R_{*}) \: R_{*},
\end{equation}
where $(R_{p} / R_{*})$ is the fitted reduced-radius parameter, and $R_{*}$ is the stellar radius. Planet radius in units of \rearth{} is displayed in Figure~\ref{f15} versus orbital period in units of days; both radius and period are plotted on logarithmic scales. The full DR25 TCE population is shown in the upper panel. The density of the radius versus orbital period distribution is shown in the middle panel. The TCEs associated with golden KOIs are displayed in the lower panel. The axes have been restricted to focus on the bulk of the population. There are a large number of TCEs with long orbital periods and relatively small planet radii. There are 1884 TCEs with period $> 300$~days and radius $< 2.0$\rearth, and 3207 TCEs with period $> 300$~days and radius $< 2.5$\rearth; these TCEs are of particular interest for the \keplermission{}, a goal of which has been to determine the frequency of Earth-sized planets in the habitable zone of Sun-like stars \citep{borucki1}. The large number of long-period TCEs associated with relatively small planets does not imply that the goal has been accomplished, however. The TCEs in this regime have high false-alarm rates and will require careful vetting to produce a reliable catalog of planetary candidates. The \kepler{} TCERT vetting activity in support of the DR25 planetary candidates catalog is currently underway.

Concerning the TCEs with orbital period $> 300$~days, the bootstrap false-alarm probability is $> 10^{-12}$ for 1025 of the 1884 TCEs (54.4\%) with radius $< 2.0$\rearth, and 1733 of the 3207 TCEs (54.0\%) with radius $< 2.5$\rearth. Taking the model fit S/N into consideration, either the bootstrap false-alarm probability is $> 10^{-12}$ or the model S/N is $< 7.1\sigma$ for 1439 of the 1884 TCEs (76.4\%) with radius $< 2.0$\rearth, and 2373 of the 3207 TCEs (74.0\%) with radius $< 2.5$\rearth. 

\begin{figure}
%%\epsscale{.80}
%\plotone{fig16.pdf}
\includegraphics[bb=34 14 510 406]{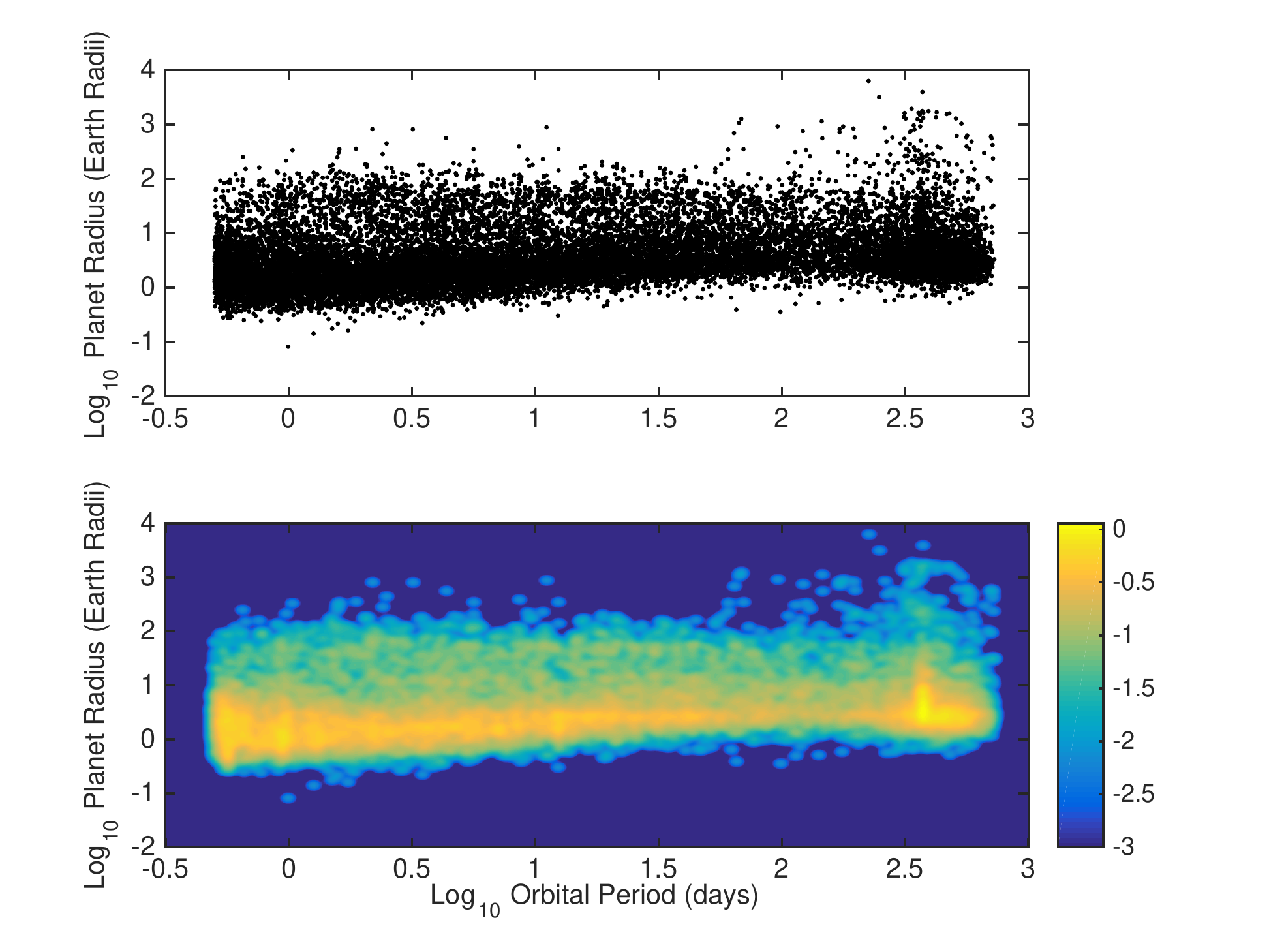}
\caption{Planet radius in \rearth{} vs. orbital period in days on logarithmic scales. Top: DR25 TCEs for which the false-alarm probability does not exceed $10^{-12}$ and the model fit S/N is not less than $7.1\sigma$. Bottom: density for TCEs in top panel on logarithmic scale.
\label{f16}}
\end{figure}

The planet radius in units of \rearth{} is displayed again in Figure~\ref{f16} versus the orbital period in units of days. The DR25 TCE population is shown in the upper panel after removal of 13,492 TCEs (39.6\% of total) with bootstrap false-alarm probability $> 10^{-12}$ or model fit S/N $< 7.1\sigma$. The density of the radius versus orbital period distribution is shown in the bottom panel. Many of the DR25 TCEs corresponding to small planets and many of the TCEs with long orbital periods do not survive the false-alarm probability and S/N cuts.

\begin{figure}
%%\epsscale{.80}
%\plotone{fig17.pdf}
\includegraphics[bb=34 16 501 393]{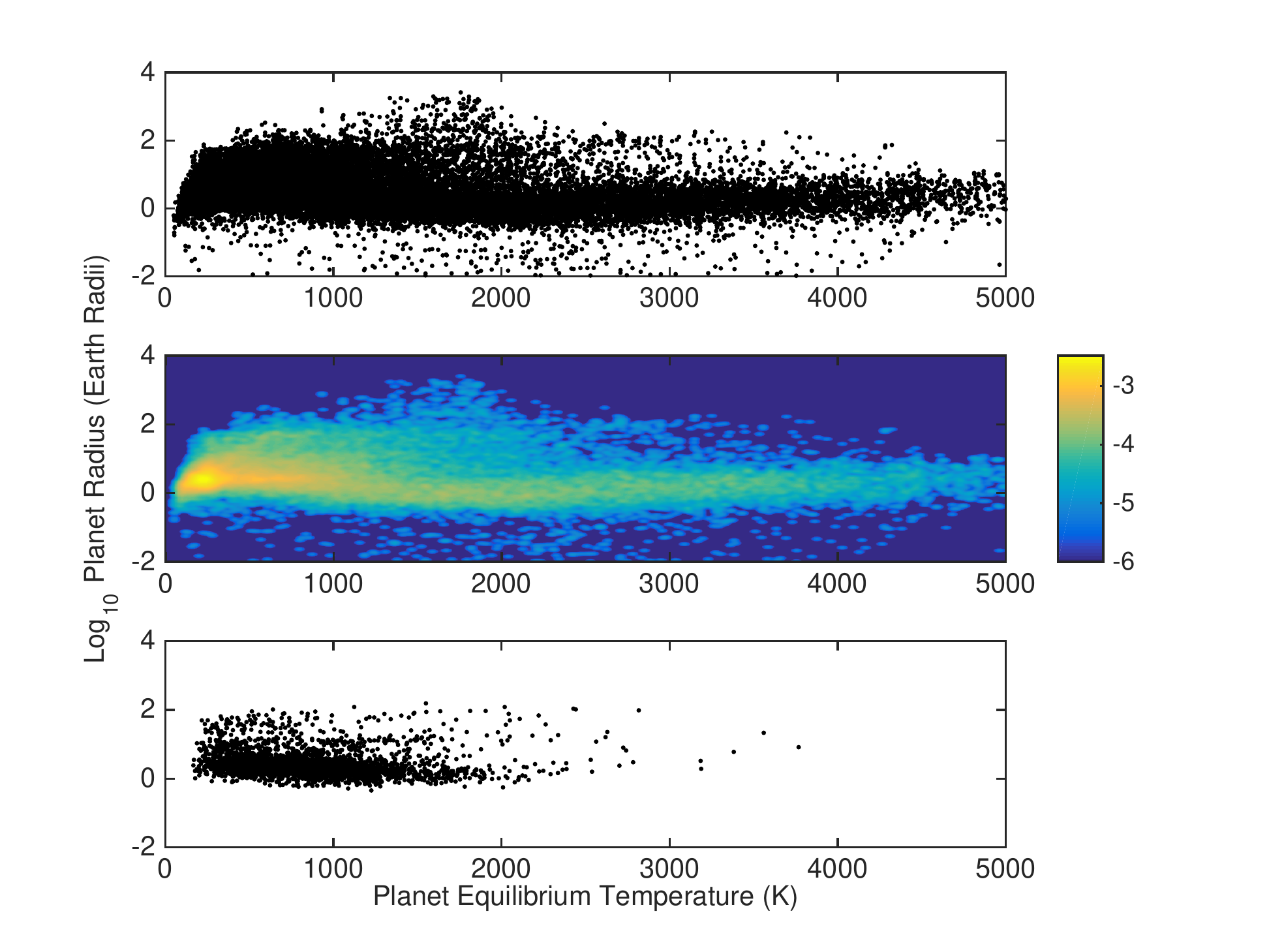}
\caption{Planet radius in \rearth{} (on logarithmic scale) vs. planet equilibrium temperature in kelvin.  Top: all DR25 TCEs. Middle: density for all TCEs on logarithmic scale. Bottom: ``golden KOIs.''
\label{f17}}
\end{figure}

We use two approaches to identify TCEs that are of particular interest to habitability studies. The first approach relies on the planet equilibrium temperature, $T_\mathrm{eq}$, which is computed for each TCE in DV by
\begin{equation}
T_\mathrm{eq} = T_{*} \: (1-\alpha)^{1/4} \: \bigg(\frac{R_{*}}{2 a_{p}}\bigg)^{1/2},
\end{equation}
where $T_{*}$ is the effective temperature of the host star, $R_{*}$ is the stellar radius, $a_{p}$ is the semimajor axis of the orbit, and $\alpha$ is the planet albedo. The albedo is assumed to be 0.3 in DV; perfect redistribution of heat is also assumed. Figure~\ref{f17} shows planet radius in units of \rearth{} versus equilibrium temperature in kelvin;  radius is plotted on a logarithmic scale. The full DR25 TCE population is shown in the upper panel. The density of the radius versus equilibrium distribution is displayed in the middle panel. The TCEs associated with golden KOIs are shown in the lower panel. The axes have been restricted to focus on the bulk of the population.

There are a large number of TCEs in the DR25 results corresponding to small, cool planets. There are 1302 TCEs with $185 < T_{eq} < 305$K and radius $< 2.0$\rearth, and 2334 TCEs with $185 < T_{eq} < 305$K and radius $< 2.5$\rearth. In this equilibrium temperature range, the bootstrap false-alarm probability is $> 10^{-12}$ for 721 of the 1302 TCEs (55.4\%) with radius $< 2.0$\rearth, and 1269 of the 2334 TCEs (54.4\%) with radius $< 2.5$\rearth. Either the bootstrap false-alarm probability is $> 10^{-12}$ or the model fit S/N is $< 7.1\sigma$ for 995 of the 1302 TCEs (76.4\%) with radius $< 2.0$\rearth, and 1712 of the 2334 TCEs (73.4\%) with radius $< 2.5$\rearth.

\begin{figure}
%%\epsscale{.80}
%\plotone{fig18.pdf}
\includegraphics[bb=34 16 510 406]{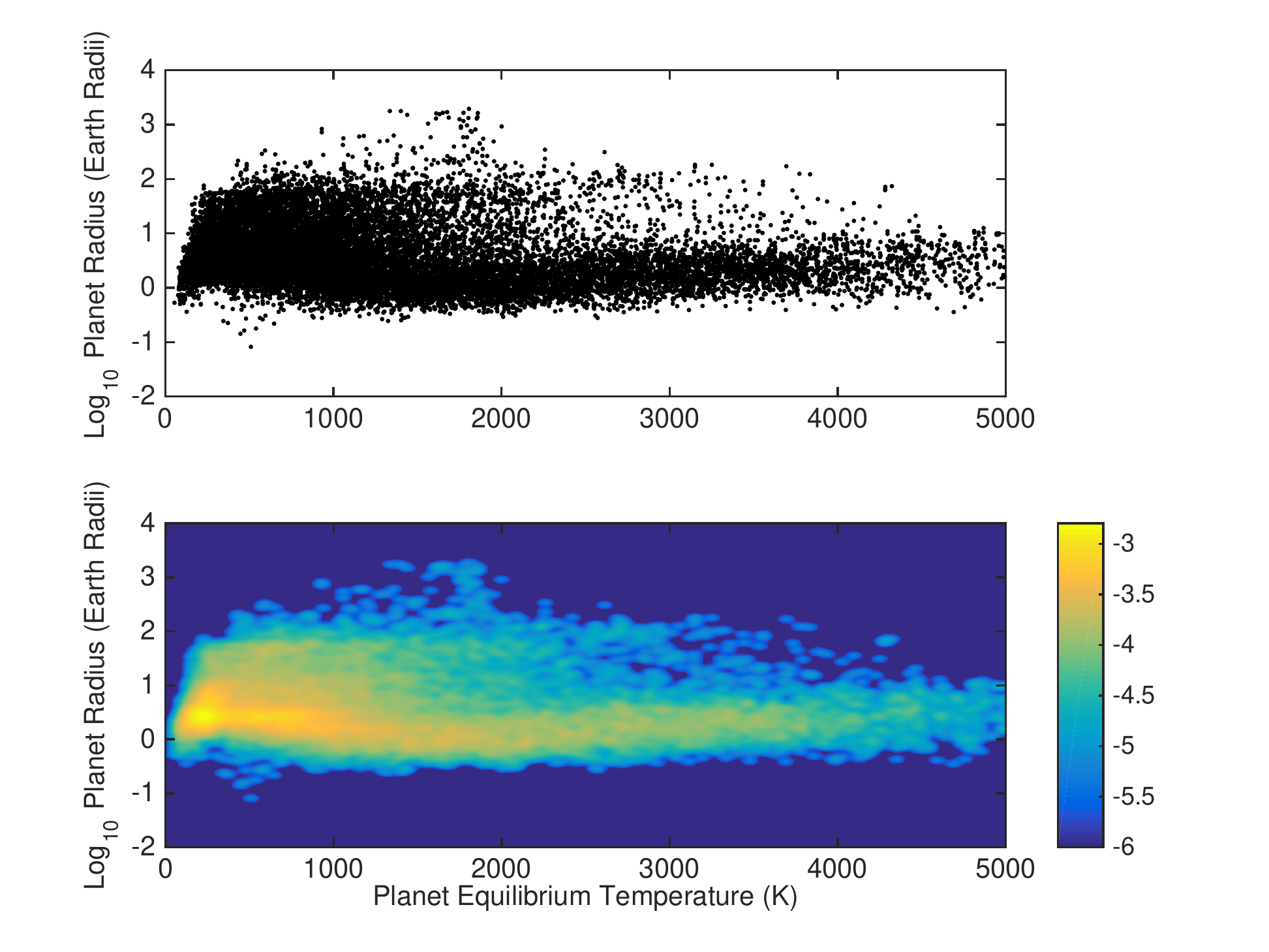}
\caption{Planet radius in \rearth{} (on logarithmic scale) vs. planet equilibrium temperature in kelvin.  Top: DR25 TCEs for which the false alarm probability does not exceed $10^{-12}$ and the model fit S/N is not less than $7.1\sigma$. Bottom: density for TCEs in top panel on logarithmic scale.
\label{f18}}
\end{figure}

Planet radius in units of \rearth{} is displayed again in Figure~\ref{f18} versus equilibrium temperature in kelvin. The DR25 TCE population is shown in the upper panel after removal of those TCEs with bootstrap false-alarm probability $> 10^{-12}$ or model fit S/N $< 7.1\sigma$. The density of the radius versus equilibrium distribution is shown in the bottom panel. Many of the DR25 TCEs corresponding to small planets and many of the TCEs corresponding to planets with relatively cool equilibrium temperatures do not survive the false-alarm probability and S/N cuts.

\begin{figure}
%%\epsscale{.80}
%\plotone{fig19.pdf}
\includegraphics[bb=34 14 510 393]{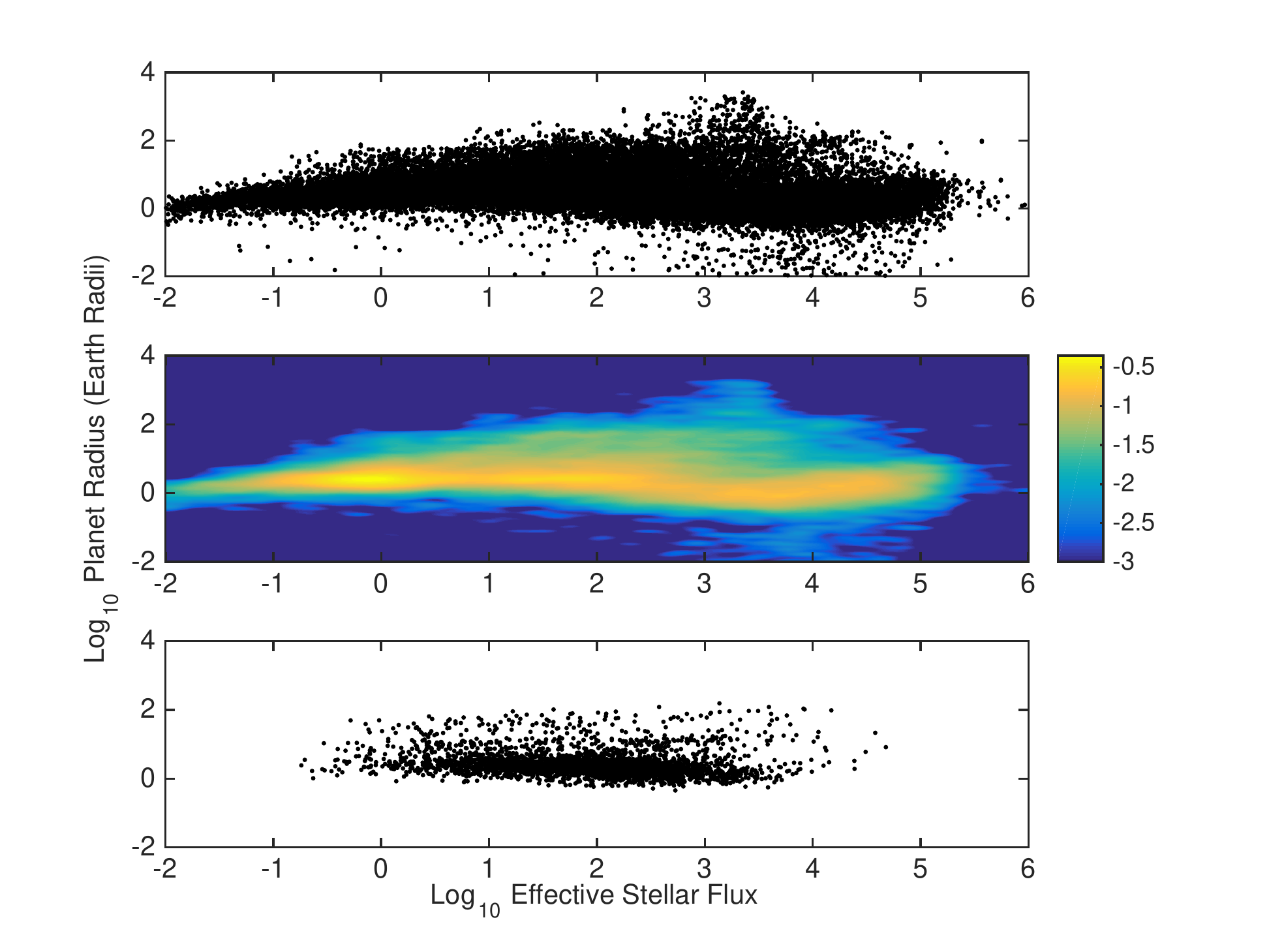}
\caption{Planet radius in \rearth{} vs. effective stellar flux on logarithmic scales.  Effective stellar flux is defined as the insolation relative to the solar flux received at the top of Earth's atmosphere. Top: all DR25 TCEs. Middle: density for all TCEs on logarithmic scale. Bottom: ``golden KOIs.''
\label{f19}}
\end{figure}

The second approach relies on the effective stellar flux, $S_\mathrm{eff}$, which is the amount of flux received from the host star at the top of a planetary atmosphere relative to the flux received from the Sun at the top of Earth's atmosphere. The relationship between effective stellar flux and the habitable zone of main-sequence stars was discussed by \citet{kopparapu2013}. The calculation of effective stellar flux requires fewer assumptions than are needed for equilibrium temperature. This parameter is derived in DV by
\begin{equation}
S_\mathrm{eff} = \bigg(\frac{R_{*}}{\rsun{}}\bigg)^{2} \: \bigg(\frac{T_{*}}{\tsun{}}\bigg)^{4} \: \bigg(\frac{a_{p}}{\aearth{}}\bigg)^{-2},
\end{equation}
where $R_{*}$ is the stellar radius, $T_{*}$ is the stellar effective temperature, and $a_{p}$ is the semimajor axis of the planet's orbit. These are normalized by $\rsun{}$ (the solar radius), $\tsun{}$ (the solar effective temperature), and $\aearth{}$ (the semimajor axis of Earth's orbit, 1 au). Figure~\ref{f19} shows planet radius in units of \rearth{} versus effective stellar flux;  radius and effective stellar flux are both plotted on logarithmic scales. The full DR25 TCE population is shown in the upper panel. The density of the radius versus effective flux distribution is displayed in the middle panel. The TCEs associated with golden KOIs are shown in the lower panel. The axes have been restricted to focus on the bulk of the population. We see again a large number of TCEs corresponding to small, cool planets.

For the record, there are 1275 TCEs with $0.25 < S_\mathrm{eff} < 1.75$ and radius $< 2.0$\rearth, and 2270 TCEs with $0.25 < S_\mathrm{eff} < 1.75$ and radius $< 2.5$\rearth. In this effective stellar flux range, the bootstrap false-alarm probability is $> 10^{-12}$ for 701 of the 1275 TCEs (55.0\%) with radius $< 2.0$\rearth, and 1224 of the 2270 TCEs (53.9\%) with radius $< 2.5$\rearth. Either the bootstrap false-alarm probability is $> 10^{-12}$ or the model fit S/N is $< 7.1\sigma$ for 970 of the 1275 TCEs (76.1\%) with radius $< 2.0$\rearth, and 1659 of the 2270 TCEs (73.1\%) with radius $< 2.5$\rearth. 

\begin{figure}
%%\epsscale{.80}
%\plotone{fig20.pdf}
\includegraphics[bb=34 14 510 406]{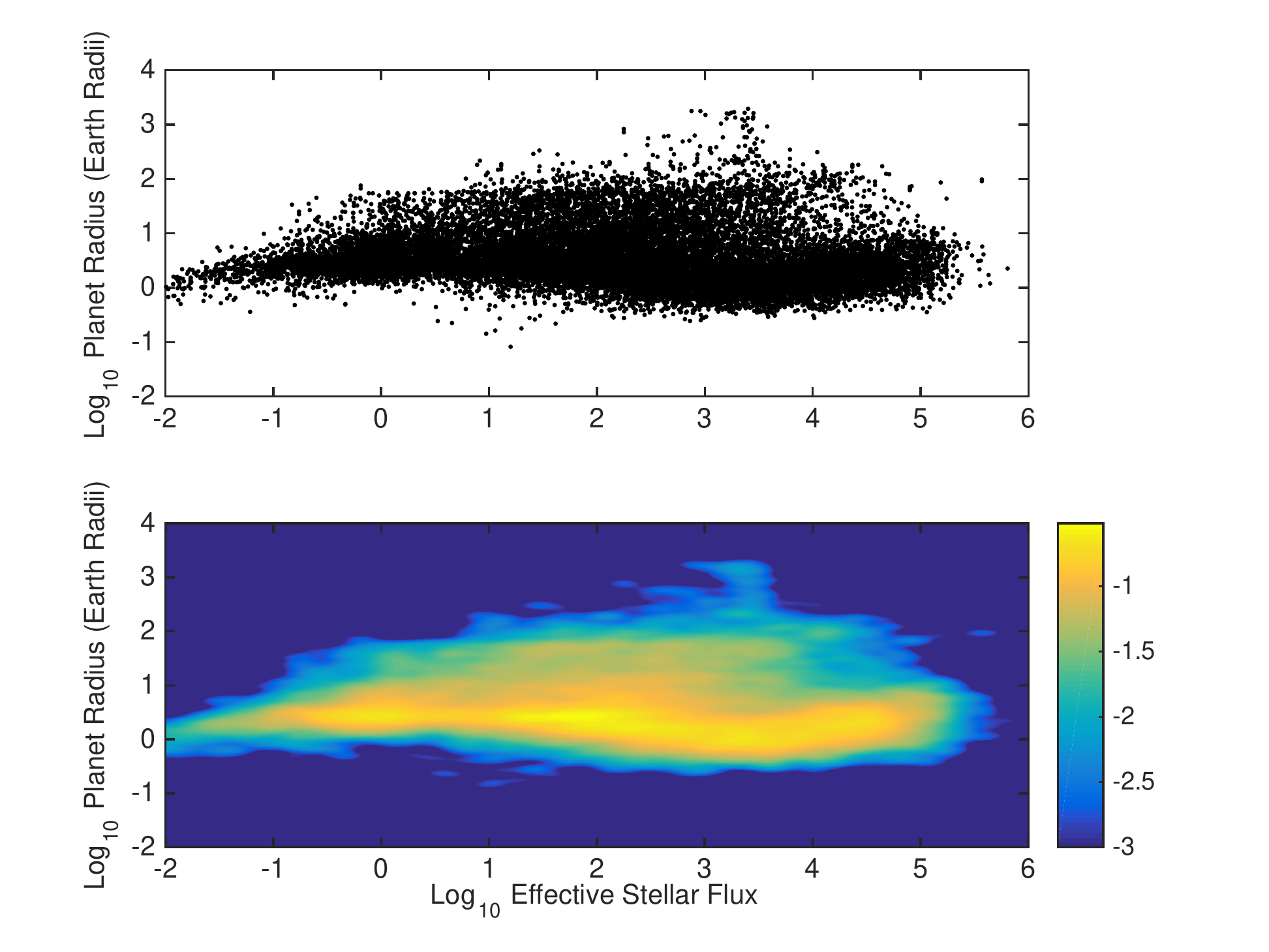}
\caption{Planet radius in \rearth{} vs. effective stellar flux on logarithmic scales.  Effective stellar flux is defined as the insolation relative to the solar flux received at the top of Earth's atmosphere. Top: DR25 TCEs for which the false-alarm probability does not exceed $10^{-12}$ and the model fit S/N is not less than $7.1\sigma$. Bottom: density for TCEs in top panel on logarithmic scale.
\label{f20}}
\end{figure}

Planet radius in units of \rearth{} is displayed again in Figure~\ref{f20} versus effective stellar flux. The DR25 TCE population is shown in the upper panel after removal of those TCEs with bootstrap false-alarm probability $> 10^{-12}$ or model fit S/N $< 7.1\sigma$. The density of the radius versus effective stellar flux is shown in the bottom panel. Many of the DR25 TCEs corresponding to small planets and many of the TCEs corresponding to planets with low insolation do not survive the false-alarm probability and S/N cuts.

\section{CONCLUSION}
\label{s:conclusion}
We present the catalog of TCEs produced in the final \kepler{} Data Processing Pipeline transiting planet search (DR25) of the full Q1--Q17 primary mission data set. The TCEs were generated by running the pipeline end-to-end with the SOC 9.3 code base. Photometric light curves of 198,709 stellar targets were searched for transit signals in a gridded space of period, epoch, and pulse duration. The search resulted in the generation of 34,032 TCEs on 17,230 unique target stars that exceeded the 7.1$\sigma$ pipeline transit detection threshold and satisfied three statistical transit signal consistency tests (i.e.,~vetoes). This result far exceeds the number of TCEs produced in earlier transit searches. The increase is due in part to increasing the sensitivity of the latest pipeline code base and operational configuration to planets with long-period transit signatures that were vetoed in the DR24 search.

The performance of the final pipeline search for transiting planets was assessed by the ability to recover a set of 3402 well-established, high-quality KOIs (``golden KOIs'') selected well in advance of the DR25 run. An analysis of the pipeline results indicates that 3394 of 3402 (99.8\%) gold-standard KOIs were recovered. That recovery rate significantly exceeds the performance of earlier \kepler{} Pipeline runs. The large number of TCEs reflects improvements in pipeline photometric precision and transiting planet search algorithms, changes to transit signal consistency tests and test criteria, and an emphasis on completeness over reliability in the final pipeline transit search. Completeness, reflected by the very high recovery rate of well-established, high-quality KOIs, must be weighed against the large number of false-alarm detections produced in this run. Pixel- and flux-level transit injection represent a better gauge of detection efficiency than recovery of established KOIs; these activities for characterizing DR25 pipeline performance are underway. 

We have shown that there are approximately 1,000--2,000 DR25 TCEs corresponding to small planets in or near the habitable zone of their host stars. The vast majority of these do not likely represent small planets orbiting in the habitable zone, however. This population of TCEs must be vetted with high accuracy to produce a final planet catalog that is reliable in the planet parameter space that is of most relevance to the \keplermission{}. The \kepler{} TCERT vetting activity in support of the DR25 planetary candidates catalog is currently underway. DR25 TCEs, model fit and validation diagnostic test results, DV Reports and Report Summaries, and DV Time Series files are available for public access at the NASA Exoplanet Archive (http://exoplanetarchive.ipac.caltech.edu).

%\section{ACKNOWLEDGMENTS}
%\label{s:ack}
\vspace{36pt}
\kepler{} was competitively selected as the 10th NASA Discovery mission. Funding for this mission is provided by the NASA Science Mission Directorate. The authors gratefully acknowledge the contributions of the greater \kepler{} team in building and operating the instrument, collecting and distributing the science data, producing the light curves and validation products employed in this publication, and archiving the results. The light curves and validation products were generated by the \kepler{} Data Processing Pipeline through the efforts of the \textit{Kepler} Science Operations Center and Science Office. The \keplermission{} is led by the project office at NASA Ames Research Center. Ball Aerospace built the \kepler{} photometer and spacecraft, which is operated by the Mission Operations Center at LASP. The light curves are archived at the Mikulski Archive for Space Telescopes; the Data Validation products are archived at the NASA Exoplanet Archive. The authors also gratefully acknowledge the support of the NASA Advanced Supercomputing (NAS) Division within the Science Mission Directorate. All of the pipeline processing described in this paper was performed on NAS Pleiades hardware; the pixel- and flux-level transit injection processing referred to in this paper was also performed on Pleiades. The authors finally wish to acknowledge the years of efforts of William J. Borucki and the late David G. Koch; the \keplermission{} would not have been possible without them.

%\begin{thebibliography}{references}
%\bibliography{references}{}

\begin{thebibliography}{}
\bibitem[Akeson et al.(2013)]{akeson} Akeson, R.~L., et al.\ 2013, \pasp, 125, 989
\bibitem[Allen(2005)]{allen} Allen, B.\ 2005, PhRvD, 71, 062001
\bibitem[Batalha et al.(2013)]{batalha1} Batalha, N.~M., et al.\ 2013, \apjs, 204, 24 
\bibitem[Borucki et al.(2010)]{borucki1} Borucki, W.~J., et al.\ 2010, Science, 327, 977
\bibitem[Borucki et al.(2011a)]{borucki2} Borucki, W.~J., et al.\ 2011a, \apj, 728, 117
\bibitem[Borucki et al.(2011b)]{borucki3} Borucki, W.~J., et al.\ 2011b, \apj, 736, 19
\bibitem[Brown et al.(2011)]{brown} Brown, T.~M., Latham, D.~W., Everett, M.~E., and Esquerdo, G.~A.\ 2011, \aj, 140, 112
\bibitem[Bryson et al.(2010)]{bryson2010} Bryson, S.~T., et al. \ 2010, Proc SPIE 7740, 77401D
\bibitem[Bryson et al.(2013)]{bryson2013} Bryson, S.~T., et al.\ 2013, \pasp, 125, 889
\bibitem[Bryson et al.(2016)]{bryson2016} Bryson, S.~T., et al.\ 2016, Kepler Certified False Positive Table (KSCI-19093-002)
\bibitem[Burke et al.(2014)]{burke1} Burke, C.~J., et al.\ 2014, \apjs, 210, 19 
\bibitem[Caldwell et al.(2010)]{caldwell} Caldwell, D.~A., et al.\ 2010, \apjl, 713, L92
\bibitem[Catanzarite(2015)]{catanzarite} Catanzarite, J.~H.\ 2015, Autovetter Planet Candidate Catalog for Q1--Q17 Data Release 24 (KSCI-19091-001)
\bibitem[Christiansen et al.(2012)]{christiansen2012} Christiansen, J.~L., et al.\ 2012, \pasp, 124, 1279
\bibitem[Christiansen et al.(2013)]{christiansen2013} Christiansen, J.~L., et al.\ 2013, \apjs, 207, 35
\bibitem[Christiansen et al.(2015)]{christiansen2015} Christiansen, J.~L., et al.\ 2015, \apj, 810, 95
\bibitem[Christiansen et al.(2016a)]{christiansen2016a} Christiansen, J.~L., et al.\ 2016a, \apj, 828, 99
\bibitem[Christiansen et al.(2016b)]{christiansen2016b} Christiansen, J.~L., et al.\ 2016b, in preparation
\bibitem[Claret \& Bloemen(2011)]{claret2011} Claret, A., \& Bloemen, S.\ 2011, \aap, 529, AA75
\bibitem[Clarke et al.(2012)]{clarke2012} Clarke, B.~D., Kolodziejczak, J.~J., and Caldwell, D.~A.\ 2012, AAS Meeting Abstracts \#220, 220, \#330.02 
\bibitem[Clarke et al.(2015)]{clarke2015} Clarke, B.~D., Caldwell, D.~A., and Kolodziejczak, J.~J.\ 2015, IAU General Assembly \#29, 2258444
\bibitem[Coughlin et al.(2014)]{jcough} Coughlin, J.~L., et al.\ 2014, \aj, 147, 119
\bibitem[Coughlin et al.(2016)]{coughlin2016} Coughlin, J.~L., et al.\ 2016, \apjs, 224, 12
\bibitem[Duda \& Hart(1972)]{hough} Duda, R.~O., \& Hart, P.~E.\ 1972, Comm. ACM, 15, 11
\bibitem[Ford et al.(2012)]{ford2012} Ford, E.~B., et al.\ 2012, \apj, 756, 185
\bibitem[Haas et al.(2010)]{science-ops} Haas, M.~R., et al.\ 2010, \apjl, 713, L115
\bibitem[Howell et al.(2014)]{howell} Howell, S.~B., et al.\ 2014, \pasp, 126, 398
\bibitem[Huber et al.(2016)]{huber2016} Huber, D., et al.\ 2016, in preparation
\bibitem[Jenkins et al.(2002a)]{jenkins2002a} Jenkins, J.~M., Caldwell, D.~A., and Borucki, W.~J.\ 2002a, \apj, 564, 495
\bibitem[Jenkins (2002b)]{jenkins2002b} Jenkins, J.~M.\ 2002b, \apj, 575, 493
\bibitem[Jenkins et al.(2010a)]{jenkins2010a} Jenkins, J.~M., et al.\ 2010a, \apjl, 713, L87 
\bibitem[Jenkins et al.(2010b)]{jmj2010b} Jenkins, J.~M., et al.\ 2010b, \procspie, 7740, 77400D
\bibitem[Jenkins et al.(2014)]{jenkins2014} Jenkins, J.~M., et al.\ 2014, Proc. IAU, 293, 94
\bibitem[Jenkins et al.(2015)]{jenkins2015} Jenkins, J.~M., et al.\ 2015, \aj, 150, 56
\bibitem[Kirk et al.(2016)]{eb-cat3} Kirk, B., et al.\ 2016, \aj, 151, 68
\bibitem[Kolodziejczak et al.(2010)]{jeffk2010} Kolodziejczak, J.~J., et al.\ 2010, \procspie, 7742, 77421G
\bibitem[Kopparapu et al.(2013)]{kopparapu2013} Kopparapu, R.~K., et al.\ 2013, \apj, 765, 131
\bibitem[Li et al.(2015)]{li2015} Li, J., et al.\ 2015, IAU General Assembly \#29, 2248472
\bibitem[Li et al.(2016)]{li2016} Li, J., et al.\ 2016, in preparation
\bibitem[Mandel \& Agol(2002)]{mandel} Mandel, K., \& Agol, E.\ 2002, \apjl, 580, L171 
\bibitem[Matijevi{\v c} et al.(2012)]{morph} Matijevi{\v c}, G., et al.\ 2012, \aj, 143, 123
\bibitem[Mazeh et al.(2013)]{mazeh2013} Mazeh, T., et al.\ 2013, \apjs, 208, 16 
\bibitem[McCauliff et al.(2015)]{mccauliff} McCauliff, S.~D., et al.\ 2015, \apj, 806, 6
\bibitem[Morris et al.(2016)]{rlm2016} Morris, R.~L., et al.\ 2016, in preparation
\bibitem[Mullally et al.(2015)]{mullally2015} Mullally, F., et al.\ 2015, \apjs, 217, 31
\bibitem[Mullally et al.(2016)]{mullally2016} Mullally, F., et al.\ 2016, \pasp, 128, 074502
\bibitem[Pr{\v s}a et al.(2011)]{eb-cat1} Pr{\v s}a, A., et al.\ 2011, \aj, 141, 83 
\bibitem[Quintana et al.(2010)]{calSpie2010} Quintana, E.~V., et al. 2010, \procspie, 7740, 77401X
\bibitem[Rowe et al.(2015)]{rowe2015} Rowe, J.~F., et al.\ 2015, \apjs, 207, 16
\bibitem[Seader et al.(2013)]{seader2013} Seader, S., Tenenbaum, P., Jenkins, J.~M., and Burke, C.~J.\ 2013, \apjs, 206, 25
\bibitem[Seader et al.(2015)]{seader2015} Seader, S., et al.\ 2015, \apjs, 217, 18 
\bibitem[Slawson et al.(2011)]{eb-cat2} Slawson, R.~W., et al.\ 2011, \aj, 142, 160 
\bibitem[Smith et al.(2012)]{js2012} Smith, J.~C., et al.\ 2012, \pasp, 124, 1000
\bibitem[Smith et al.(2016)]{js2016} Smith, J.~C., et al.\ 2016, \pasp, 128, 124501
\bibitem[Stumpe et al.(2012)]{stumpe2012} Stumpe, M.~C., et al.\ 2012, \pasp, 124, 985 
\bibitem[Stumpe et al.(2014)]{stumpe2014} Stumpe, M.~C., et al.\ 2014, \pasp, 126, 100
\bibitem[Tenenbaum et al.(2010)]{pt2010} Tenenbaum, P., et al.\ 2010, \procspie, 7740, 77400J
\bibitem[Tenenbaum et al.(2012)]{pt2012} Tenenbaum, P., et al.\ 2012, \apjs, 199, 24 
\bibitem[Tenenbaum et al.(2013)]{pt2013} Tenenbaum, P., et al.\ 2013, \apjs, 206, 5 
\bibitem[Tenenbaum et al.(2014)]{pt2014} Tenenbaum, P., et al.\ 2014, \apjs, 211, 6
\bibitem[Thompson et al.(2015a)]{thompson2015a} Thompson, S.~E., et al.\ 2015a, \apj, 812, 46
\bibitem[Thompson et al.(2015b)]{thompson2015b} Thompson, S.~E., et al.\ 2015b, Kepler Data Release 25 Notes (KSCI-19065-001)
\bibitem[Thompson(2016)]{thompson2016} Thompson, S.~E.\ 2016, Data Validation Time Series File: Description of Format and Content (KSCI-19079-001)
\bibitem[Twicken et al.(2010)]{jdt2010} Twicken, J.~D., et al.\ 2010, \procspie, 7740, 77401U
\bibitem[Twicken et al.(2014)]{jdt2014} Twicken, J.~D., et al.\ 2014, AAS Meeting Abstracts \#224, 224, \#120.06
\bibitem[Twicken et al.(2015)]{jdt2015} Twicken, J.~D., et al.\ 2015, IAU General Assembly \#29, 2252899
\bibitem[Twicken et al.(2016)]{jdt2016} Twicken, J.~D., et al.\ 2016, in preparation
\bibitem[Van Cleve et al.(2016)]{vancleve} Van Cleve, J.~E., et al.\ 2016, \pasp, 128, 075002
\bibitem[Wu et al.(2010)]{hw2010} Wu, H., et al.\ 2010, \procspie, 7740, 774019
\end{thebibliography}
%\bibliographystyle{abbrv}

% There were issues rendering PDF images at ApJ and arXiv due to lack of bounding boxes. The bb coordinates may be obtained as follows:
%
%    $ gs -o /dev/null -sDEVICE=bbox fig6.pdf
%    GPL Ghostscript 9.16 (2015-03-30)
%    Page 1
%   %%BoundingBox: 38 14 510 394
%   %%HiResBoundingBox: 38.537999 14.778000 509.651984 393.767988
%
% The figure may then be loaded by:
%
%    \includegraphics[bb=38 14 510 394]{fig6.pdf}

\clearpage
\onecolumn
\clearpage
\clearpage
\clearpage
\clearpage
\clearpage
\clearpage
\clearpage
\clearpage
\clearpage
\clearpage
\clearpage
\clearpage
\clearpage
\clearpage
\clearpage
\clearpage
\clearpage
\clearpage
\clearpage

\clearpage

%% The values (usually only l,r and c) in the last part of
%% \begin{deluxetable}{} command tell LaTeX how many columns
%% there are and how to align them.
\begin{deluxetable}{ccc}

%% Keep a portrait orientation

%% Over-ride the default font size
%% Use Default (12pt)

%% Use \tablewidth{?pt} to over-ride the default table width.
%% If you are unhappy with the default look at the end of the
%% *.log file to see what the default was set at before adjusting
%% this value.
\tablewidth{280pt}

%% This is the title of the table.
\tablecaption{Targets Searched for Transiting Planets\label{t1}}

%% This command over-rides LaTeX's natural table count
%% and replaces it with this number.  LaTeX will increment 
%% all other tables after this table based on this number
\tablenum{1}

%% The \tablehead gives provides the column headers.  It
%% is currently set up so that the column labels are on the
%% top line and the units surrounded by ()s are in the 
%% bottom line.  You may add more header information by writing
%% another line between these lines. For each column that requires
%% extra information be sure to include a \colhead{text} command
%% and remember to end any extra lines with \\ and include the 
%% correct number of &s.

\tablehead{\colhead{KIC ID} & \colhead{Number of} & \colhead{Number of} \\
\colhead{} & \colhead{Quarters Observed} & \colhead{TCEs} \\} 

%% All data must appear between the \startdata and \enddata commands
\startdata
757076 &   17 &    0\\
757099 &   17 &    1\\
757137 &   17 &    0\\
757280 &   17 &    0\\
757450 &   17 &    1\\
891901 &   17 &    0\\
891916 &   17 &    0\\
892010 &     3 &    0\\
892107 &   17 &    0\\
892195 &   17 &    0\\
892203 &   17 &    0\\
892376 &   17 &    6\\
892667 &   17 &    1\\
892675 &   17 &    0\\
892678 &   17 &    0\\
892713 &   17 &    0\\
892718 &   14 &    0\\
892738 &   17 &    0\\
892760 &     5 &    0\\
892772 &   14 &    1\\
\enddata

\bigskip
(This table is available in its entirety in machine-readable form.)

%% Include any \tablenotetext{key}{text}, \tablerefs{ref list},
%% or \tablecomments{text} between the \enddata and 
%% \end{deluxetable} commands

%% No \tablecomments indicated

%% No \tablerefs indicated

\end{deluxetable}

\clearpage

%% The values (usually only l,r and c) in the last part of
%% \begin{deluxetable}{} command tell LaTeX how many columns
%% there are and how to align them.
\begin{deluxetable}{c}

%% Keep a portrait orientation

%% Over-ride the default font size
%% Use Default (12pt)

%% Use \tablewidth{?pt} to over-ride the default table width.
%% If you are unhappy with the default look at the end of the
%% *.log file to see what the default was set at before adjusting
%% this value.
\tablewidth{60pt}

%% This is the title of the table.
\tablecaption{Eclipsing Binaries Excluded from Transiting Planet Search\label{t2}}

%% This command over-rides LaTeX's natural table count
%% and replaces it with this number.  LaTeX will increment 
%% all other tables after this table based on this number
\tablenum{2}

%% The \tablehead gives provides the column headers.  It
%% is currently set up so that the column labels are on the
%% top line and the units surrounded by ()s are in the 
%% bottom line.  You may add more header information by writing
%% another line between these lines. For each column that requires
%% extra information be sure to include a \colhead{text} command
%% and remember to end any extra lines with \\ and include the 
%% correct number of &s.

\tablehead{\colhead{KIC ID}  \\} 

%% All data must appear between the \startdata and \enddata commands
\startdata
1433410 \\
1572353 \\
1868650 \\
2012362 \\
2141697 \\
2159783 \\
2162283 \\
2302092 \\
2305277 \\
2435971 \\
2437038 \\
2444187 \\
2448320 \\
2449084 \\
2450566 \\
2453212 \\
2570289 \\
2571439 \\
2577756 \\
2694741 \\
\enddata

\bigskip
(This table is available in its entirety in machine-readable form.)

%% Include any \tablenotetext{key}{text}, \tablerefs{ref list},
%% or \tablecomments{text} between the \enddata and 
%% \end{deluxetable} commands

%% No \tablecomments indicated

%% No \tablerefs indicated

\end{deluxetable}

\clearpage

%% The values (usually only l,r and c) in the last part of
%% \begin{deluxetable}{} command tell LaTeX how many columns
%% there are and how to align them.
\begin{deluxetable}{cccccccc}

%% Keep a portrait orientation

%% Over-ride the default font size
%% Use Default (12pt)

%% Use \tablewidth{?pt} to over-ride the default table width.
%% If you are unhappy with the default look at the end of the
%% *.log file to see what the default was set at before adjusting
%% this value.

%% This is the title of the table.
\tablecaption{Golden KOI and TCE Ephemeris Matching Results\label{t3}}

%% This command over-rides LaTeX's natural table count
%% and replaces it with this number.  LaTeX will increment 
%% all other tables after this table based on this number
\tablenum{3}

%% The \tablehead gives provides the column headers.  It
%% is currently set up so that the column labels are on the
%% top line and the units surrounded by ()s are in the 
%% bottom line.  You may add more header information by writing
%% another line between these lines. For each column that requires
%% extra information be sure to include a \colhead{text} command
%% and remember to end any extra lines with \\ and include the 
%% correct number of &s.

\tablehead{\colhead{KOI} & \colhead{KOI} & \colhead{TCE} & \colhead{KOI} & \colhead{TCE} & \colhead{KOI} & \colhead{TCE} & \colhead{Correlation} \\
\colhead{Number} & \colhead{Period} & \colhead{Period} & \colhead{Epoch} & \colhead{Epoch} & \colhead{Duration}  & \colhead{Duration} & \colhead{Coefficient} \\
\colhead{} & \colhead{(days)} & \colhead{(days)} & \colhead{(BKJD)} & \colhead{(BKJD)} & \colhead{(hr)} & \colhead{(hr)} & \colhead{} } 

%% All data must appear between the \startdata and \enddata commands
\startdata
1.01 & 2.4706 & 2.4706 & 122.7633 & 132.6457 & 1.7426 & 1.7968 & 0.985 \\ 
2.01 & 2.2047 & 2.2047 & 121.3586 & 132.3833 & 3.8822 & 4.0438 & 0.980 \\ 
3.01 & 4.8878 & 4.8878 & 124.8131 & 134.5888 & 2.3639 & 2.4059 & 0.991 \\ 
4.01 & 3.8494 & 3.8494 & 157.5267 & 134.4299 & 2.6605 & 2.7133 & 0.990 \\ 
5.01 & 4.7803 & 4.7803 & 132.9741 & 132.9740 & 2.0349 & 2.0611 & 0.994 \\ 
7.01 & 3.2137 & 3.2137 & 123.6119 & 133.2543 & 3.9935 & 4.1268 & 0.984 \\ 
10.01 & 3.5225 & 3.5225 & 121.1194 & 131.6870 & 3.1906 & 3.2711 & 0.988 \\ 
12.01 & 17.8552 & 17.8552 & 146.5964 & 146.5961 & 7.4294 & 7.4400 & 0.999 \\ 
13.01 & 1.7636 & 1.7636 & 120.5659 & 132.9115 & 3.1814 & 3.1796 & 0.998 \\ 
17.01 & 3.2347 & 3.2347 & 121.4866 & 134.4253 & 3.6011 & 3.5910 & 0.999 \\ 
18.01 & 3.5485 & 3.5485 & 122.9015 & 133.5470 & 4.5770 & 4.5615 & 0.998 \\ 
20.01 & 4.4380 & 4.4380 & 171.0091 & 135.5055 & 4.7010 & 4.6918 & 0.999 \\ 
22.01 & 7.8914 & 7.8914 & 177.2500 & 137.7928 & 4.3040 & 4.3537 & 0.994 \\ 
41.01 & 12.8159 & 12.8159 & 122.9482 & 135.7632 & 6.3728 & 6.5001 & 0.990 \\ 
41.02 & 6.8871 & 6.8871 & 133.1779 & 133.1786 & 4.4276 & 4.5472 & 0.987 \\ 
41.03 & 35.3331 & 35.3331 & 153.9833 & 153.9855 & 5.9040 & 6.0418 & 0.982 \\ 
42.01 & 17.8337 & 17.8338 & 181.2337 & 145.5634 & 4.5403 & 4.8219 & 0.970 \\ 
46.01 & 3.4877 & 3.4877 & 170.9320 & 132.5683 & 3.8427 & 3.9487 & 0.987 \\ 
46.02 & 6.0298 & 6.0297 & 132.4818 & 132.4809 & 3.9100 & 4.0465 & 0.952 \\ 
49.01 & 8.3138 & 8.3138 & 175.9916 & 134.4234 & 2.9927 & 3.1174 & 0.980 \\
\enddata

\bigskip
(This table is available in its entirety in machine-readable form.)

%% Include any \tablenotetext{key}{text}, \tablerefs{ref list},
%% or \tablecomments{text} between the \enddata and 
%% \end{deluxetable} commands

%% No \tablecomments indicated

%% No \tablerefs indicated

\end{deluxetable}

\end{document}